\documentclass{aa}  

\usepackage{graphicx}
\usepackage{booktabs}
\usepackage{subfigure}
\usepackage{mathrsfs}
\usepackage{txfonts}
\usepackage[colorlinks=true, citecolor=blue, linkcolor = blue]{hyperref}
\usepackage{gensymb}

\newcommand{\dd}{\ensuremath{\mathrm{d}}}
\newcommand{\msafit}{{\sc msafit}}
\newcommand{\ppxf}{{\sc ppxf}}
\newcommand{\vobsi}{$V_{i,{\rm obs}}$}
\newcommand{\sigmaobsi}{$\sigma_{v,i,{\rm obs}}$}
\newcommand{\kms}{km~s$^{-1}$}

\begin{document}

   \title{Jeans Dynamical Constraints on the IMF in Massive Quiescent Galaxies at $z\sim0.7$ from JWST/NIRSpec}

   \author{Martje Slob\inst{1}\corrauth{slob@strw.leidenuniv.nl}
          \and 
          Mariska Kriek\inst{1}
          \and
          Chloe M. Cheng\inst{1,2}
          \and 
          Arjen van der Wel\inst{3}
          \and
          Aliza G. Beverage\inst{4}
          \and
          Rachel Bezanson\inst{5}
          \and
          Charlie Conroy\inst{6}
          \and
          Anna de Graaff\inst{6,7}
          \and
          Elham Eftekhari\inst{1}
          \and
          Robert Feldmann\inst{8}
          \and
          Wout M. Goesaert\inst{1}
          \and
          Meng Gu\inst{9,10}
          \and
          Joel Leja\inst{11,12,13}
          \and
          Brian Lorenz\inst{14}
          \and
          Pavel E. Mancera Pi\~na\inst{1}
          \and
          Ignacio Mart\'in-Navarro \inst{15,16}
          \and
          Andrew B. Newman\inst{4}
          \and
          Sedona H. Price\inst{17}
          \and
          Alice E. Shapley\inst{18}
          \and
          Katherine A. Suess\inst{19}
          \and 
          Jesse van de Sande\inst{20}
          \and
          Pieter van Dokkum\inst{21}
          }

   \institute{Leiden Observatory, Leiden University, P.O. Box 9513, 2300 RA Leiden, The Netherlands.  
   \and
   David A. Dunlap Department of Astronomy and Astrophysics, University of Toronto, Toronto, ON M5S 3H4, Canada
   \and
    Sterrenkundig Observatorium, Universiteit Ghent, Krijgslaan 281 S9, B-9000 Gent, Belgium
   \and 
   Observatories of the Carnegie Institution for Science, 813 Santa Barbara St., Pasadena, CA 91101, USA
   \and
   Department of Physics and Astronomy and PITT PACC, University of Pittsburgh, Pittsburgh, PA 15260, USA
   \and
   Center for Astrophysics $|$ Harvard \& Smithsonian, Cambridge, MA 02138, USA
   \and
   Max-Planck-Institut f\"ur Astronomie, K\"onigstuhl 17, D-69117, Heidelberg, Germany
   \and
   Department of Astrophysics, Universit\"at Z\"urich, Zurich CH-8057, Switzerland
   \and
   Department of Astronomy, Tsinghua University, Beijing 100084, People's Republic of China
    \and
    Hong Kong Institute for Astronomy \& Astrophysics, Pokfulam Road, The University of Hong Kong, Pok Fu Lam, Hong Kong
    \and
    Department of Astronomy \& Astrophysics, The Pennsylvania State University, University Park, PA 16802, USA
    \and
    Institute for Computational \& Data Sciences, The Pennsylvania State University, University Park, PA 16802, USA
    \and
    Institute for Gravitation and the Cosmos, The Pennsylvania State University,  University Park, PA 16802, USA
    \and
    Department of Astronomy, University of California, Berkeley, CA 94720, USA
    \and 
    Instituto de Astrof\'isica de Canarias,  Vía Láctea, E38205  La Laguna, Tenerife, Spain
    \and
    Departamento de Astrof\'isica, Universidad de La Laguna, E-38205 La Laguna, Tenerife, Spain
    \and 
    Space Telescope Science Institute (STScI), 3700 San Martin Drive, Baltimore, MD 21218, USA
    \and
    Department of Physics \& Astronomy, University of California, Los Angeles, 430 Portola Plaza, Los Angeles, CA 90095, 
    \and
    Department for Astrophysical \& Planetary Science, University of Colorado, Boulder, CO 80309, USA
    \and
    School of Physics, University of New South Wales, Sydney, NSW 2052, Australia
    \and
    Astronomy Department, Yale University, 52 Hillhouse Ave, New Haven, CT 06511, USA
   }
   \date{Received XX; accepted XX}
 
  \abstract{
  We present dynamical masses for nine massive quiescent galaxies at $z\sim0.7$ from the JWST-IMFERNO program, measured from ultra-deep ($\sim32$~hr), spatially resolved NIRSpec/MSA spectroscopy combined with NIRCam imaging. The most massive of these galaxies have previously been found to have a bottom-heavy initial mass function (IMF), based on measurements of dwarf-sensitive absorption features. Here, we examine whether the stellar masses inferred for these IMFs are consistent with the dynamical masses of these galaxies. We derive the dynamical masses by including Jeans Anisotropic Modelling within a forward-modelling framework that accounts for instrumental optics, source morphology, and geometric effects. We compare the resulting dynamical masses to the stellar masses for different IMF assumptions. A Milky Way IMF produces physically plausible dynamical-to-stellar mass ratios but implies central dark matter fractions ($f_{\rm DM,1R_e}=0.54\pm0.06$) that are a factor $\sim2$ higher compared to local ETGs and simulations. When uniformly applying the IMF that was measured from absorption features, the mean dark matter fraction becomes low ($f_{\rm DM,1R_e}=0.02\pm0.18$), and the stellar masses exceed the dynamical mass at $2\sigma$ significance for two out of seven galaxies in the sample. When we adopt the measured IMF only in the central $R_e$ and use a MW IMF in the outskirts, we find realistic dark matter fractions ($f_{\rm DM,1R_e}=0.28\pm0.12$) and consistent dynamical masses. These results support a scenario where the cores of the most massive quiescent galaxies at $z\sim0.7$ had a bottom-heavy IMF, similar to those observed in the centres of local early-type galaxies. Our results are consistent with an inside-out growth scenario, in which size and mass growth of massive galaxies proceeds after quenching through minor mergers, leaving the bottom-heavy cores mostly intact.
  }


   \maketitle
   \nolinenumbers

\section{Introduction}\label{sec:intro}
In recent years, the James Webb Space Telescope (JWST) has revealed an increasing number of extremely massive quiescent galaxies at $z>3$ \citep[e.g.][]{AdeGraaff2024_highz_q,KGlazebrook2023,ACarnall2024}, challenging our models of galaxy formation \citep{JAntwiDanso2025,WBaker2025,AWeibel2025,AdeGraaff2024_highz_q, CLagos2025}.
This tension may be even more severe than initially recognized, as the stellar masses of these galaxies could have been systematically underestimated due to assumptions about the stellar initial mass function (IMF). Studies generally adopt a universal IMF calibrated to the Milky Way (MW) disk. However, \citet{CCheng2026} showed that the likely descendants of these early quiescent galaxies may instead have bottom-heavy IMFs, increasing their inferred stellar masses by factors of up to four relative to a MW-like IMF. If similarly bottom-heavy IMFs were already in place at high redshift, the inferred stellar masses of the population of massive quiescent galaxies at $z>4$ would increase substantially, making them even more difficult to reconcile with galaxy formation models.

Although \citet{CCheng2026} provide the first direct IMF measurements beyond the local universe, evidence for a bottom-heavy IMF in massive early-type galaxies (ETGs) is well established at $z\sim0$ \citep[e.g.][]{TTreu2010, CSpiniello2011, JThomas2011, ADutton2012, CConroy2012d, MCappellari2012, cappellari_atlas3d_2013, ANewman2013, ANewman2017, MGu2022, PvanDokkum2010_IMF, PvanDokkum2012_IMF, conroy_slar_2012}. Spatially resolved studies further show that this bottom-heavy IMF is concentrated in galaxy centres, while the outskirts approach a MW IMF \citep{IMartinNavarro2015,FLaBarbera2016,TDavis2017,PvanDokkum2017,MKianusch2024, MGu2026}.
These central regions of local ETGs are thought to be the descendants of the compact quiescent galaxies at $z\gtrsim2$ \citep{TNaab2009, RBezanson2009, PvanDokkum2010, RFeldmann2010, oser_two_2010, JvandeSande2013}, supporting a picture in which high-$z$ massive quiescent galaxies have a bottom-heavy IMF. Furthermore, relic galaxies, which are proposed to be the undisturbed descendants of high-$z$ massive quiescent galaxies \citep[e.g.][]{trujillo_ngc_2014,AFerreMateu2017} are found to have bottom-heavy IMFs at all radii \citep{IMartinNavarro2015, martin-navarro_inspire_2023, maksymowicz-maciata_inspire_2024}.

Although a bottom-heavy IMF at $z\sim0.7$ is not unexpected, the \citet{CCheng2026} results rely on sensitive spectroscopic features, and require independent confirmation. One avenue for such confirmation is provided by dynamical mass measurements. \citet{CCheng2026} showed that their inferred stellar masses are supported by virial masses primarily derived from deep VLT/VIMOS rest-frame optical spectra from the LEGA-C survey \citep{AvanderWel2021}. However, virial masses are highly dependent on assumptions of structural parameters, slit losses, and rotational support \citep[e.g.][]{MCappellari2006,SBelli2017,AvanderWel2022,MKriek2024,MSlob2025}, and were measured using seeing limited ground-based observations. More accurate dynamical constraints based on high spatial resolution spectroscopy are therefore required to independently confirm whether this bottom-heavy IMF is consistent with the dynamical masses.

Such robust dynamical masses of quiescent galaxies from stellar kinematics can be obtained from spatially resolved two-dimensional integral field spectroscopy (IFS), dynamically modelled through e.g. Jeans Anisotropic Modelling (JAM; \citealt{MCappellari2008, MCappellari2020}). Ground-based surveys such as MASSIVE \citep{CMa2014,MVeale2017}, SAMI \citep{SCroom2012,SCroom2021,JvandeSande2017} and ATLAS3D \citep{MCappellari2011,EEmsellem2011}, enable such analyses for large samples at $z\sim0$, while \citet{mozumdar_magnus_2026} extend this work to intermediate redshifts using VLT/MUSE observations. However, beyond $z\gtrsim0.5$ detailed IFS observations remain limited by the extreme integration times required to measure kinematics from faint stellar absorption features, as well as the fact that currently available instruments can only observe one target at a time. Such observations have thus far only been obtained for a handful of distant quiescent galaxies \citep{FDEugenio2023, RPascalau2025, forrest_massive_2025, ANewman2025}.

An efficient alternative to the IFU is provided by the micro-shutter array (MSA) on JWST/NIRSpec \citep{PFerruit2022}. This observing mode can obtain slit-based spectra of $\sim70$ galaxies simultaneously, and has a high sensitivity and spatial resolution. Although the observations only cover one spatial dimension, randomly orientated with respect to the galaxy major axis, the intrinsic kinematics can be recovered through a forward-modelling framework that combines the resolved 2D spectra with morphological constraints from imaging and detailed models of instrumental effects. This forward modelling method has already been developed for simple rotating disk models of gas \citep{AdeGraaff2023} and stellar kinematics (\citealt{MSlob2025}, see also \citealt{SPrice2016, SPrice2020, JvanHoudt2021, CStraatman2022}).

In this work, we implement JAM modelling within such a forward-modelling framework to analyse the stellar kinematics of NIRSpec/MSA observations of the JWST-IMFERNO sample. Such JAM modelling is necessary to obtain reliable dynamical masses from NIRSpec/MSA data as it provides a more realistic description of the stellar kinematics compared to the simple rotating disc models used in previous works. Furthermore, while a similar method has been developed for ground-based VLT/VIMOS observations of the full LEGA-C sample from which JWST-IMFERNO was drawn \citep{JvanHoudt2021}, the higher spatial resolution of both the spectroscopic and imaging JWST data used in this work allows us to extend this work to more compact galaxies and provides better constraints on the dynamical masses for all galaxies.

Combined with the IMF measurements from \citet{CCheng2026}, this work provides the first opportunity beyond the local universe to directly compare IMF constraints from stellar absorption line modelling with independent dynamical mass measurements in the same systems. The galaxies were initially drawn from the LEGA-C survey \citep{AvanderWel2021}, located in the COSMOS field, and observed in a single NIRSpec/MSA pointing. High-resolution JWST/NIRCam imaging from the COSMOS-Web program \citep{CCasey2023} is available for all targets, enabling our JAM-based forward-modeling approach.

The paper is organised as follows. In Section \ref{sec:data} we present the spectroscopic and imaging data used in this work. Our forward modelling technique and stellar mass modelling approach are outlined in Section \ref{sec:methods}. We present our dynamical modelling results and dynamical masses in Section \ref{sec:results}. In Section \ref{sec:discussion} we discuss the implications for the IMF and galaxy evolution of our results, as well as possible caveats in our work. Finally, in Section \ref{sec:conclusion} we summarise our findings. Throughout this work we assume a $\Lambda$ cold dark matter cosmology with $\Omega_m = 0.3, \Omega_{\Lambda}=0.7$ and $H_0 = 70$~\kms~Mpc$^{-1}$.

\section{Sample and Data}\label{sec:data}
\begin{figure*}[t]
    \centering
    \includegraphics[width=\linewidth]{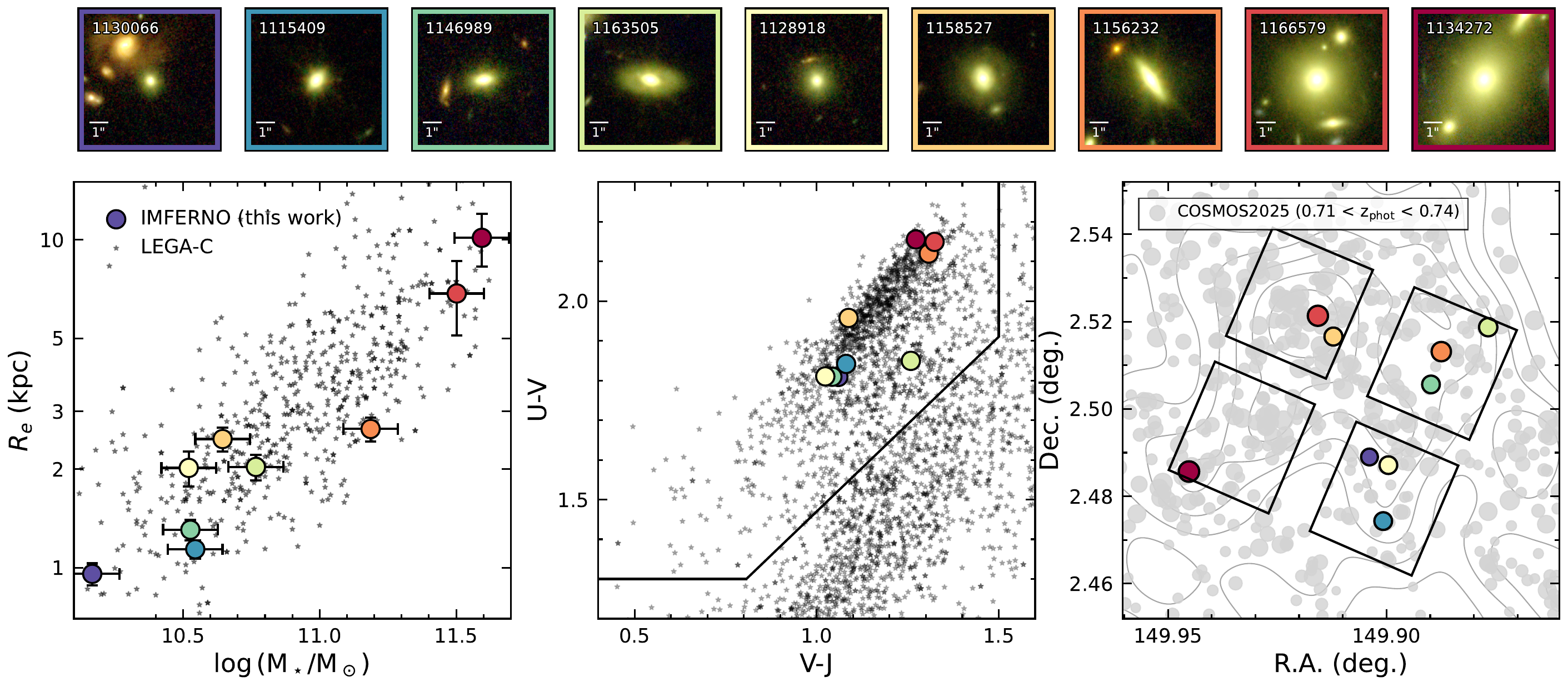}
    \caption{Top panels: 8''$\times$8'' NIRCam RGB cutouts from COSMOS-Web imaging \citep{CCasey2023} for the galaxies in our sample, ordered by size. The colours of the panels correspond to the points in the bottom panels. Bottom left panel: stellar mass vs. size in NIRCam F150W for the galaxies in our sample (coloured points). We also show mass vs. HST F160W sizes for the population of LEGA-C galaxies from which our sample was selected (black stars). Bottom centre panel: rest-frame $U-V$ vs. $V-J$ colours for our sample (coloured points), and the general population of LEGA-C galaxies (black stars). We indicate the quiescent galaxy selection box, defined by \citet{AMuzzin2013b}. Bottom right panel: Location in the sky of the galaxies in our sample, with the NIRSpec/MSA footprint shown in black. We show all galaxies in COSMOS2025 with $0.71<z_{\rm phot}<0.74$ as gray points, with the corresponding density contours in the background. The relative sizes of the COSMOS2025 markers are proportional to their stellar mass.}
    \label{fig:sample_overview}
\end{figure*}
This study is based on observations from the NIRSpec/MSA JWST-IMFERNO programme (JWST-GO-5629, \citealt{CCheng2026}), which observed thirteen massive, quiescent galaxies at $z\sim0.7$. The galaxies in the JWST-IMFERNO sample were selected from the optical VLT/VIMOS spectroscopic LEGA-C survey \citep{AvanderWel2021} to be quiescent and sufficiently bright ($H$-band magnitude $<21.5$), and to cover IMF sensitive features at rest-frame $\sim8100-9950$~\AA. Below, we provide a summary of key steps in the observation and reduction of the NIRSpec/MSA spectra. For a detailed description of the target selection and observation/reduction strategy we refer to \citet{CCheng2026}.

We observed our targets with the G140M-F100LP disperser and filter combination on the NIRSpec/MSA, with a total on-source integration time of 31.2 hours. This grating/filter combination has a medium spectral resolution ($R\sim 1500$), and covers rest-frame $\sim5600-10800$~\AA\ for our sample. We used a custom two-point nodding pattern, with a nod size of two shutters (1.06\arcsec), to ensure that the extended galaxies in our sample are not self-subtracted in the nodded background subtraction. Furthermore, our MSA slits cover $4-8$ shutters to capture (nearly) the full spatial extent of our galaxies. To correct for detector artifacts we also include a three shutter dither in the dispersion direction.
The raw data for both of these dithers were reduced using version 1.20.2 (CRDS version 1464) of the JWST Science calibration software \citep{HBushouse2023} to obtain combined, rectified, flux calibrated 2D spectra.

To obtain the spatially resolved spectral information needed to measure the dynamical properties of our sample, we perform the following steps (see also \citealt{MSlob2025}). First, we extract a 1D spectrum for each individual row of the reduced 2D spectra, which is then corrected for a sinusoidal pattern (``wiggles''), caused by the undersampling of the NIRSpec point spread function (PSF) along the spatial direction of the 2D spectra. The wiggle correction is done following the method described in \citet{CCheng2025_gradients} (see also \citealt{MPerna2023}), by fitting a sinusoidal function to the spectrum of each row, and dividing this pattern out of the data. Finally, for each row, we combine the wiggle-corrected 1D spectra from the two dithered positions to obtain the row-by-row spectra for the galaxies in our sample.\\

To obtain light distributions of the targets in our sample, detailed rest-frame optical imaging is needed. To this end, we use imaging from the COSMOS-Web programme (PID 1727, \citealt{CCasey2023}) in the F115W, F150W, F277W and F444W filters. For our analysis, we use the imaging data available in MAST, which were reduced using the JWST Calibration Pipeline v1.17.14 with the CRDS pipeline mapping (pmap) context 1230.\\

In Figure \ref{fig:sample_overview} we show overviews of the galaxies in our sample in the mass-size plane (see Section \ref{sec:met_mgefit}), $UVJ$ space, and the footprint of the survey, as well as RGB images of our sample. For reference, we also show the quiescent galaxies in the LEGA-C sample from which our sample was selected. We adopt the COSMOS2020 \citep{JWeaver2022} IDs of the targets in our sample throughout this work, following \citet{CCheng2026}.

\section{Methods}\label{sec:methods}
Accurate dynamical mass measurements of distant quiescent galaxies require resolved dynamical modelling that properly accounts for rotational support and projection effects. Achieving this with NIRSpec/MSA observations is non-trivial, as the data sample only a single spatial dimension with a randomly orientated orientation and placement within the slit. However, using a detailed forward modelling approach that includes measurements of a galaxy's position and orientation with respect to the MSA slits, we can recover the intrinsic kinematic properties of galaxies from their 2D NIRSpec/MSA spectra \citep{AdeGraaff2023, MSlob2025, ARavishankar2026}. 

\begin{figure*}[t]
    \centering
    \includegraphics[width=0.9\linewidth]{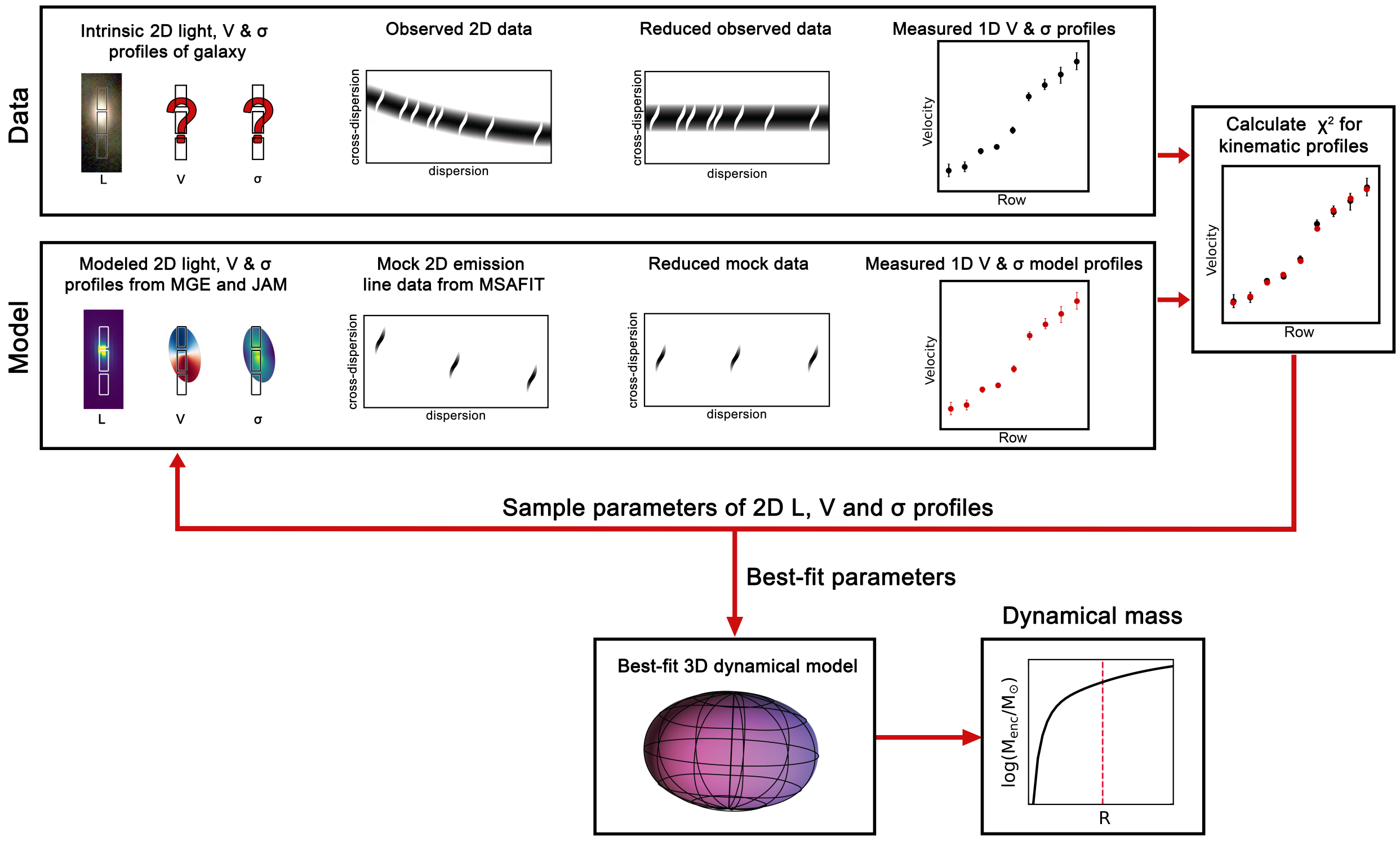}
    \caption{Graphic representation of the method used to infer the best-fit dynamical mass.}
    \label{fig:flowchart}
\end{figure*}

In this study, we expand on the parametric stellar kinematic forward modelling approach presented in \citet{MSlob2025}, by implementing detailed dynamical JAM modelling to constrain stellar dynamics. In this framework the galaxy's dynamical structure is modelled from the light distribution with respect to the MSA slit, to obtain a description of the intrinsic three-dimensional mass distribution. We describe the steps in our forward modelling procedure below, and provide an illustration of the key steps in Figure \ref{fig:flowchart}.
\subsection{Kinematic measurements}\label{sec:met_v_mods}
We measure the resolved observed stellar velocity and dispersion profiles of the galaxies in our sample by fitting the stellar velocity dispersion ($\sigma_{v,i,\text{obs}}$) and velocity shift ($V_{i,\text{obs}}$) for each extracted row $i$ (see Section \ref{sec:data}) using the fitting code \ppxf\ \citep{MCappellari2004, MCappellari2017, MCappellari2023}, and the e-MILES spectral library \citep{AVazdekis2016}. To account for the instrumental resolution, we use the JDox line spread function curve\footnote{\url{https://jwst-docs.stsci.edu/jwst-near-infrared-spectrograph/nirspec-instrumentation/nirspec-dispersers-and-filters}}, rescaled with a constant factor derived using MSAFIT for the morphology of each individual galaxy, following \citet{MSlob2024}. The currently implemented models in \ppxf struggle to model the continuum shape of the full wavelength range of our spectra, leading to poor fits to the spectra. The polynomial order we consequently required to obtain a robust fit to the full wavelength range is of the order $\sim50$, which resulted in high uncertainties in our kinematic fits.  Thus, to prevent template mismatches affecting our kinematic measurements and obtain kinematic measurements for the largest number of spectral rows possible, we only use the (rest-frame) wavelength range $8400-8800$~\AA, covering the Calcium triplet feature. 
Nonetheless, we find that fitting the full wavelength range does not lead to systematically different results, with a mean offset between the measured kinematics per row of $6.76\pm3.4$~km s$^{-1}$.

We restrict the stellar population templates used in \ppxf\ in the resolved analysis, by first fitting the integrated spectrum of the galaxy. To correct for any flux calibration issues we include a 2nd-order additive polynomial in this fit. The resulting stellar population template is then used to fit the \sigmaobsi\ and \vobsi\ of each row $i$ with sufficiently high $S/N$ ($S/N>4$~pixel$^{-1}$; \citealt{JvandeSande2017, FDEugenio2023}). We calculate the errors on \sigmaobsi\ and \vobsi\ by applying the same fitting algorithm to 1000 Monte Carlo simulations of the spectrum, which were constructed by perturbing the spectrum using the error spectrum. To ensure that our velocity (dispersion) measurements are reliable, we only include rows for which the error on \vobsi\ or \sigmaobsi\ is smaller than 70\,km/s. For larger errors, we find that the models do not fit the data well. Finally, in order to constrain our dynamical models, we require a galaxy to have kinematic data in at least 5 rows. Our selection requirements result in a final sample of 9 galaxies for which we can model the stellar kinematics, of which 7 are in common with \citet{CCheng2026}.

\subsection{Dynamical Modelling}\label{sec:dynamical_models}
To interpret the resolved stellar velocity (dispersion) data measured in the previous section we use dynamical models based on the JAM framework \citep{MCappellari2008}, implemented using the \texttt{JAMpy} package\footnote{\url{https://pypi.org/project/jampy/}}. In this approach we solve the hydrodynamic Jeans equations to recover the dynamical properties of a galaxy, assuming axisymmetry. We use a description of the gravitational potential derived from the observed surface brightness distribution and a parametrised dark matter (DM) component. In JAM, the gravitational potential is parametrised as a set of 2D Gaussians through a multi-Gaussian expansion (MGE; \citealt{MCappellari2002}) of both the light distribution and dark matter halo. Below, we describe how the MGE of the gravitational potential was constructed, and how we set up the JAM models.

\subsubsection{Light models}\label{sec:met_mgefit}
To model the intrinsic surface brightness of the galaxies in our sample, we first fit a double S\'ersic model to all four observed NIRCam photometric bands using \texttt{GALFIT} \citep{CPeng2002} (see Appendix \ref{ap:galfit} for a detailed description of the fitting procedure). Overall, the best-fit parameters of our double S\'ersic models agree well between the different photometric filters. In the rest of this work we use the surface brightness profiles from the F150W bands, as this filter most closely matches the wavelength coverage of our NIRSpec/MSA data.

Next, we perform an MGE fit of the PSF de-convolved double S\'ersic model image. This step results in a set of elliptical Gaussians that describe the surface brightness profile, where each Gaussian $j$ is parametrised by the peak flux $F_j$, the size of the major axis $\sigma_j$, and the axial ratio $q_j$. To ensure we obtain a stable solution in our JAM modelling we restrict the axial ratios of the Gaussians to the axis ratio of the galaxy from Table \ref{tab:results}. The final step in generating a description of the surface brightness is to convert the peak flux of the Gaussians into units of $L_{\odot}~{\rm pc}^{-2}$, as required by the \texttt{JAMpy} package. To this end, we first calculate the total rest-frame luminosity of our galaxy in the SDSS $z$ filter (roughly corresponding to the same rest-frame filter as NIRCam/F150W). We then convert our peak flux values to units of $L_{\odot}~{\rm pixel}^{-2}$ using the total solar luminosity in the same filter, calculated with the \citet{GThuillier2003} solar spectrum. Finally, we convert from pixel$^{-2}$ to pc$^{-2}$ using the pixel scale of our images and the angular scale at the redshift of the galaxy. 

Our MGE fits describe the S\'ersic line profiles well, with a typical mean absolute deviation between the MGE fit and S\'ersic profile of $\sim1\%$. Furthermore, the GALFIT fits to the NIRCam imaging provide good models of the observed light profiles, with a typical reduced $\chi^2 \sim 0.7$. Taken together, our light models provide an accurate description of the surface brightness distributions of the galaxies in our sample.

\subsubsection{Mass models}\label{sec:met_mass_models}
In our JAM description, we model the gravitational potential using two components. The first component represents the stellar mass distribution, and follows directly from the MGE description of the surface brightness profile, described in the previous section. We convert the surface brightness profile to a surface potential profile by scaling the peak fluxes of the MGEs by the stellar mass-to-light ratio, $M_{\star}/L_\star$, which is a free parameter in our fitting procedure. 

The second component in the surface potential traditionally represents the dark matter contribution, and is likewise described using a set of 2D Gaussians. However, given the limited spatial extent and resolution of our data, we are unable to constrain the shape of the DM halo. Instead, this DM component mainly acts as a way to introduce radial gradients in the $M/L$ ratio within the region probed by our data (see also \citealt{JvanHoudt2021}). Although we parametrise the second component using a conventional DM halo model, in practice it captures any deviation from a strict mass-follows-light model. Such deviations may arise from dark matter, but could also reflect gas contributions, IMF variations or gradients in the stellar population.

We parametrise the DM density profile using a spherically symmetric NFW profile \citep{JNavarro1997}, described as 
\begin{equation}
\rho(r) = \frac{\Delta_c \rho_{\text{crit}}}{(r/r_s)(1+r/r_s)^2}.
\end{equation}
Here $r_s$ is the characteristic scale, $\rho_{\rm crit}$ the critical density, and $\Delta_c$ the threshold level delimiting the halo, conventionally chosen to be 200. In order to simplify our DM description we define the concentration parameter $c_{200} = r_{200}/r_s$, and the halo mass $M_{200}$ inside radius $r_{200}$ for which the average density is 200$\rho_{\rm crit}$. Following \citet{JvanHoudt2021} we can then eliminate $c_{200}$ from our DM description, using the correlation between halo mass, concentration and redshift. To this end, we use the relations inferred from simulations presented by \citet{ADutton2014} (though note there is scatter around this relation; e.g. \citealt{BDiemer2015});
\begin{equation}
    \log_{10}(c_{200}) = a + b \log_{10}(M_{200} / [10^{12} h^{-1}M_{\odot}]),
\end{equation}
with
\begin{align}
    a &= 0.520 + (0.905 - 0.520)e^{-0.167 z^{1.21}} \\
    b &= -0.101 + 0.026 z.
\end{align}
We can thus describe our DM component using only the overall normalisation $M_{200}$, which is a free parameter in our fitting procedure. We parametrise this DM density profile as a series of spherical ($q_j = 1$) Gaussians and add these to the stellar MGE terms. The resulting set of MGEs, described using $M_{200}$ and $M_*/L_*$, parametrises the surface potential used in our JAM modelling procedure.

Naturally, by eliminating $c_{200}$ from the fit parameters, the DM contribution is constrained to have a fixed shape. However, over the narrow radial range ($\sim 1R_e$) and limited spatial resolution ($\sim 0.1"$) probed by our observations, different halo profiles and stellar orbital anisotropies can produce similar enclosed mass profiles \citep[e.g.][]{gerhard_line--sight_1993,lokas_dark_2003,MCappellari2008, courteau_galaxy_2014}. We therefore do not attempt to constrain the detailed shape of the dark matter halo from our observations.
We also note again that the modelled DM component should not be interpreted as a constraint on the DM halo mass, but instead represents the total deviation from a strict mass-follows-light assumption. We discuss the implications of this assumption in Section \ref{sec:disc_caveats}.

\subsubsection{JAM models}\label{sec:met_JAM}
With the description of the surface density potential from the previous section, we now use JAM \citep{MCappellari2008} to generate models of the line-of-sight velocity and velocity dispersion. In our modelling, we use cylindrical coordinates, making the assumption that the velocity ellipsoid is aligned with the main cylindrical axis. We introduce the anisotropy parameter ($\beta_z = 1 - \left<v_z^2\right>/\left<v_r^2\right>$) to model the flattening of the velocity ellipsoid in the direction perpendicular to the radial plane. Using these assumptions, we can solve the Jeans equations to obtain a 3D dynamical model of our galaxy. Finally, we use the inclination, $i$, of the galaxy to project the 3D dynamics into a line-of-sight profile.

Generally, with JAM modelling the second velocity moment $(V_{\rm rms, ~los} = \sqrt{V^2_{\rm \star, ~los} +\sigma^2_{\rm \star, ~los}})$ is calculated. However, for our forward modelling set-up (see Section \ref{sec:met_msafit}) we require the second velocity moment to be decomposed into a velocity ($V_{\rm \star, ~los}$) and velocity dispersion ($\sigma_{\rm \star, ~los}$) component. To this end, we introduce the fitting parameter $\kappa$, which allows us to also generate a model for the line-of-sight first velocity moment ($V_{\star,\rm ~los}$) by scaling the velocity contribution to $V_{\rm rms}$. Following \citet{MCappellari2008}, we define $\kappa$ such that for $\kappa = 0$ there is no rotation, while for $\kappa=1$ the rotational ellipsoid is a circle in the ($v_r, v_{\phi}$) plane. We can then decompose the second velocity moment into a velocity and velocity dispersion contribution using $\sigma_{\star,\rm ~los} = \sqrt{V_{\rm rms,~ los}^2 - V_{\rm \star, ~los}^2}$. 

\subsection{Mock spectral models}\label{sec:met_msafit}
With the stellar light profile and intrinsic kinematics of the galaxies specified in the previous sections, the next step in our forward modelling procedure is to account for the optical effects of NIRSpec/MSA. We generate spectral line profiles using \msafit\ \citep{AdeGraaff2023}, which simulates the PSF, optics, and detector response given a source's morphology and position within an MSA slit.  We refer to \citet{AdeGraaff2023} for a detailed description of the standard version of \msafit, and provide a summary and our modifications to \msafit\ below.

In its standard implementation \msafit\ describes the light profile, velocity, and velocity dispersion fields using analytic parametrisations. To incorporate JAM modelling, we modify the code to allow for user-provided two-dimensional maps of the line-of-sight light profile, velocity, and velocity dispersion fields. These input maps must account for the position and alignment of a source within the MSA slit. We therefore take the 2D light profile from \texttt{MGEFIT} (see Section \ref{sec:met_mgefit}) and the projected $V_{\rm \star, ~los}$ and $\sigma_{\star,\rm ~los}$ fields from JAM (see Section \ref{sec:met_JAM}), and shift and rotate them to match the observed MSA slit position and position angle. To reduce computational cost during PSF convolution, the maps are subsequently downsampled to the pixel scale of the \msafit\ PSF model. The resulting light, velocity and velocity dispersion maps are used as the input models for \msafit.

The model spectral line profiles are generated from the input maps following the methods of \citet{MSlob2025}. We simulate three emission lines, evenly spaced across the wavelength range of the disperser/filter combination. While a larger number of lines would decrease model uncertainties by more finely sampling the detector grid, the computational costs increases steeply with the number of modelled lines. Hence, the three lines are a compromise between computation time and model error. For each line, we construct a model flux cube ($I(x,y,\lambda)$) from the input light, velocity and velocity dispersion maps. This cube is convolved with a 2$\times$ oversampled NIRSpec/MSA PSF model, generated following \citet{AdeGraaff2023}, and projected onto the detector model, accounting for the trace corresponding to the slit in which the galaxy was observed ($s_{i,j}$). The resulting detector models correspond to noiseless, unrectified single-frame observations, and include slit losses, bar shadows, spatial undersampling and interpixel capacitance. 

To match the models to the observation setup, we generate a mock spectrum for both nod positions in our observations ($s_{i,j}$ and $s_{i,j-2}$). These mock observations are processed using the same reduction steps applied to the data (Section \ref{sec:data}) and combined accordingly. The resulting model spectra therefore provide direct counterparts to the observed 2D spectra used to derive the kinematic profiles in Section \ref{sec:met_v_mods}.

Finally, we extract 1D kinematics from our models by fitting a Gaussian profile to each spectral row for all three emission lines. The median across the lines is then taken at each spectral row to obtain the final line-of-sight velocity and velocity dispersion profiles.

\subsection{Parameter inference}\label{sec:met_par_inference}
Our full dynamical model depends on seven parameters; the five parameters ($\beta, ~\log_{10}(M_\star/L), ~\kappa, ~i, ~\log_{10}(M_{200})$) described in Section \ref{sec:dynamical_models}, and two additional parameters ($\dd x, ~\dd y$) which describe the fractional offset of the galaxy position in the slit relative to the nominal pointing. We add these two parameters to account for potential pointing or astrometry inaccuracies in our observations. 
The morphological parameters are fixed to the best-fit \texttt{GALFIT} values when constructing the light profiles, as their uncertainties are small and do not significantly affect the inferred dynamical masses (see Section \ref{sec:disc_caveats}). 

We marginalise over the seven parameters using the \texttt{scipy} implementation of the differential evolution algorithm \citep{RStorn1997}, evaluated using the $\chi^2$ metric. This approach efficiently samples the full parameter space and is robust against degeneracies and complex likelihood structures, while requiring substantially fewer model evaluations than MCMC methods, which are computationally unfeasible for our methods.
The best-fit solution from differential evolution is subsequently refined using a local least-squares minimisation.

\begin{table}[t]
    \caption{Free parameters in the fit}
    \centering
    \renewcommand{\arraystretch}{1.1}
    \begin{tabular}{l l l}
        \hline Parameter & Sampling bounds & Units \\\hline
         $\beta_z$ & [-1.5, 1] & --\\
         $\log_{10}(M_\star/L_\star)$ & [-3, 2] & $\log_{10}(M_{\odot}/L_{\odot})$ \\
         $i$ & [$i_{\rm init}-20, i_{\rm init}+20$] & \textdegree \\
         $\kappa$ & [0,1] & --\\
         $\dd x$ & [-0.25, 0.25] & Shutter width\\
         $\dd y$ & [-0.25, 0.25] & Shutter length\\
         $\log_{10}(M_{200})$ & [5, 15] & $\log_{10}(M_{\odot})$\\\hline
    \end{tabular}
    \label{tab:fit_params}
\end{table}

\begin{table*}[t]
    \renewcommand{\arraystretch}{1.1}
    \caption{GALFIT measurements and JAM modelling results}
    \centering
    \vspace{-0.1in}
    \begin{tabular}{l *{9}{l}}
    \hline\hline
        \multicolumn{5}{c}{} & \multicolumn{2}{c}{{Structural parameters\tablefootmark{a}}} & \multicolumn{3}{c}{{Inferred JAM model parameters}}\\
        \cmidrule(lr){6-7}\cmidrule(lr){8-10}
        ID & $z_{\rm spec}$ & $\alpha_{\rm IMF}$\tablefootmark{b}  & $\log_{10}(M_{\star,\rm MW})$ & $\log_{10}(M_{\rm vir})$ & $r_{e, \rm tot}$\tablefootmark{c} & $q$\tablefootmark{c} & $i$ & $\beta_z$ & $\log_{10}(M_{\rm JAM})$\\
         & & & ($M_{\odot}$) & ($M_{\odot}$) & (kpc) & & (\textdegree) & & ($M_{\odot}$) \\\hline
        1115409 & 0.730 & ${-}$& ${10.5}_{-0.1}^{+0.1}$& ${10.8}_{-0.1}^{+0.1}$ & 1.14 $\pm$ 0.07 & 0.69 $\pm$ 0.08 &${36.6}_{-1.2}^{+18.1}$ &${-0.9}_{-0.3}^{+1.5}$ &${10.95}_{-0.15}^{+0.10}$\\
        1128918 & 0.715 & ${-}$& ${10.5}_{-0.1}^{+0.1}$& ${10.6}_{-0.1}^{+0.1}$ & 2.02 $\pm$ 0.24 & 0.85 $\pm$ 0.03 &${20.3}_{-4.2}^{+17.0}$ &${-0.1}_{-0.7}^{+0.3}$ &${10.78}_{-0.07}^{+0.10}$\\
        1130066 & 0.735 & ${1.37}_{-0.51}^{+0.95}$& ${10.2}_{-0.1}^{+0.1}$& ${10.3}_{-0.1}^{+0.1}$ & 0.96 $\pm$ 0.07 & 0.82 $\pm$ 0.03 &${41.0}_{-0.4}^{+1.9}$ &${0.8}_{-0.4}^{+0.1}$ &${10.45}_{-0.09}^{+0.05}$\\
        1134272 & 0.734 & ${4.01}_{-0.85}^{+0.71}$& ${11.6}_{-0.1}^{+0.1}$& ${12.0}_{-0.1}^{+0.1}$ & 10.13 $\pm$ 1.85 & 0.76 $\pm$ 0.04 &${34.4}_{-0.3}^{+0.7}$ &${-0.7}_{-0.4}^{+0.3}$ &${12.19}_{-0.06}^{+0.04}$\\
        1146989 & 0.732 & ${1.14}_{-0.36}^{+0.62}$& ${10.5}_{-0.1}^{+0.1}$& ${10.7}_{-0.1}^{+0.1}$ & 1.31 $\pm$ 0.10 & 0.45 $\pm$ 0.00 &${88.5}_{-1.1}^{+3.8}$ &${-0.5}_{-0.5}^{+0.7}$ &${11.10}_{-0.08}^{+0.05}$\\
        1156232 & 0.735 & ${2.43}_{-0.60}^{+0.75}$& ${11.2}_{-0.1}^{+0.1}$& ${11.4}_{-0.1}^{+0.1}$ & 2.65 $\pm$ 0.22 & 0.43 $\pm$ 0.03 &${68.2}_{-5.5}^{+27.1}$ &${0.2}_{-0.2}^{+0.3}$ &${11.30}_{-0.04}^{+0.04}$\\
        1158527 & 0.728 & ${2.60}_{-0.98}^{+1.21}$& ${10.6}_{-0.1}^{+0.1}$& ${11.0}_{-0.1}^{+0.1}$ & 2.47 $\pm$ 0.20 & 0.81 $\pm$ 0.04 &${41.4}_{-14.7}^{+6.4}$ &${-0.0}_{-0.1}^{+0.0}$ &${10.76}_{-0.06}^{+0.06}$\\
        1163505 & 0.730 & ${1.63}_{-0.51}^{+1.00}$& ${10.8}_{-0.1}^{+0.1}$& ${10.9}_{-0.1}^{+0.1}$ & 2.03 $\pm$ 0.18 & 0.54 $\pm$ 0.02 &${66.8}_{-7.5}^{+8.8}$ &${-0.5}_{-0.2}^{+0.3}$ &${11.16}_{-0.05}^{+0.04}$\\
        1166579 & 0.728 & ${1.39}_{-0.34}^{+0.51}$& ${11.5}_{-0.1}^{+0.1}$& ${11.8}_{-0.1}^{+0.1}$ & 6.86 $\pm$ 1.75 & 0.90 $\pm$ 0.01 &${19.5}_{-0.6}^{+1.8}$ &${-0.7}_{-0.4}^{+0.6}$ &${12.07}_{-0.08}^{+0.10}$\\
    \hline\hline
    \end{tabular}\\
    \tablefoot{
        \tablefoottext{a}{Errors on the structural parameters are calculated using the error on the fit and the scatter between the four NIRCam filters measured.}\\
        \tablefoottext{b}{IMF mismatch parameter from \citet{CCheng2026}.}
        \tablefoottext{c}{The structural parameters reported are the light-weighted mean of the two S\'ersic models fit to the imaging (see Section \ref{sec:met_mgefit}).}
    }
    \vspace{-0.25in}
    \label{tab:results}
\end{table*}

As required by the differential evolution algorithm, we adopt uniform priors for all free parameters, with bounds listed in Table \ref{tab:fit_params}. The sampling is intialised with 10 independent populations, each randomly drawn from the fitting ranges.

We obtain uncertainties on the best-fit parameters, models, and inferred dynamical masses by creating 5000 Monte Carlo simulations of the data (\vobsi\ and \sigmaobsi), each randomly perturbed around the uncertainties of the data points. We impose a noise floor of 20~\kms\ on our datapoints to avoid underestimating errors. For each realisation, the best-fit model is selected from the set of models evaluated during the differential evolution sampling. Final parameters and model uncertainties were then computed from the 16th and 84th percentiles of all realisations. We use these best-fit models and parameters to compute the dynamical masses, described in Section \ref{sec:res_dyn_masses}. 

\subsection{Stellar masses}\label{sec:met_stellar_masses}
We derive stellar masses for the galaxies in our sample using the stellar population fitting code \texttt{prospector v2.0}\footnote{\url{https://github.com/bd-j/prospector/tree/main}} \citep{JLeja2019_b,BJohnson2021} by modelling the NIRSpec/MSA IMFERNO and LEGA-C spectra, and photometry simultaneously. We use the COSMOS2025 photometric catalog \citep{MShuntov2025} for our photometric measurements. The catalog covers 37 bands spanning 0.3~$\mu$m to 8~$\mu$m, including NIRCam and MIRI imaging from COSMOS-Web \citep{CCasey2023}. We adopt the S\'ersic model fluxes in each filter to ensure we are not biased by aperture effects. We use the Flexible Stellar Population Synthesis (\texttt{FSPS}; \citealt{CConroy2009, CConroy2010}) stellar population models, with the MILES spectral library \citep{JFalcon2011} and the MIST isochrones \citep{JChoi2016, ADotter2016}, adopting a \citet{PKroupa2001} IMF. 

We adopted the \citet{MKriek2013} dust law, and fixed the birth cloud depth such that young stars are twice as attenuated as old stars. We assumed the parametric delay-$\tau$ SFH model, where we fit for the age and time scale of the model. We also leave the stellar mass, metallicity, and velocity dispersion as free parameters in the model. The redshifts are fixed to the redshifts reported in \citet{CCheng2026}. We include a 10th-order polynomial for both the LEGA-C and IMFERNO spectra, to correct for any potential flux calibration mismatches between the stellar templates and the observed spectra. Finally, we include a noise model through a spectroscopic jitter term adopted from \citet{BJohnson2021} and a pixel outlier model. We account for the spectral resolution of the LEGA-C and IMFERNO spectra separately, which were derived by \citet{CCheng2026}. Finally, we mask out any strong emission lines and the NaI D absorption feature. We present the free parameters and prior ranges of our modelling in Appendix \ref{ap:prospector}.

\section{Results}\label{sec:results}
In this section, we present the dynamical properties and masses of nine massive quiescent galaxies at $z\sim0.7$, as derived from modelling their ultra-deep JWST/NIRSpec MSA spectra. We present the results of our morphological, dynamical, and SED fitting in Table \ref{tab:results}. In Figure \ref{fig:overviews} we show the measured and best-fit line-of-sight stellar velocity and velocity dispersion profiles -- resulting from our dynamical modelling -- in 1D (row-by-row) and unconvolved projected 2D space (bottom panels) for each galaxy. We also show the COSMOS-Web RGB images, constructed from the F115W, F277W and F444W filters, with the positions of the MSA shutters overlaid on the light, velocity, and velocity dispersion maps (top panels).

For all nine galaxies in our sample the best-fit models reproduce the observed 1D velocity and velocity dispersion profiles well, with a mean reduced $\chi_{\nu}^2 = 1.07$. We therefore conclude that the best-fit three-dimensional dynamical models provide a robust description of the underlying gravitational potential of the galaxies in our sample. In Appendix \ref{ap:posteriors} we show an example corner plot for a representative galaxy.

\subsection{Stellar masses}
We report the best-fit stellar masses from our \texttt{prospector} fits in Table \ref{tab:results}.
The stellar masses we derive are consistent (mean offset $=-0.03\pm0.05$~dex) with those adopted by \citet{CCheng2026}, which were based on LEGA-C spectra, ULTRAVista photometry \citet{AMuzzin2013a}, and derived by \citet{AdeGraaff2021} using MAGPHYS \citep{EdaCunha2008MNRAS.388.1595D} for a MW IMF. Thus, the different fitting codes used and the inclusion of photometric and spectroscopic data from JWST has not led to systematic differences between the stellar masses used in this work and those adopted by \citet{CCheng2026}. Furthermore, we tested how our stellar masses change if we fix the stellar metallicity to the best-fit value from \citet{CCheng2026} instead of leaving the stellar metallicity free in the fit, and find they are in good agreement, with a mean offset of 0.002 dex. Regardless, stellar mass estimates from SED fitting are subject to systematic uncertainties arising from modelling assumptions, including the adopted elemental abundance pattern, dust properties, and model uncertainties \citep{CConroy2009, marchesini_evolution_2009, choi_imprint_2019,jones_impact_2025, conroy_modeling_2013, JLeja2019, lower_how_2020, CCheng2025_UVJ, SBellstedt2025}. As these uncertainties are not captured by the formal MCMC error budget, we adopt a minimum noise floor of 0.1~dex on our stellar mass measurements to approximately account for them.

\begin{figure*}[t]
    \centering
    \subfigure{\includegraphics[width=0.30\textwidth]{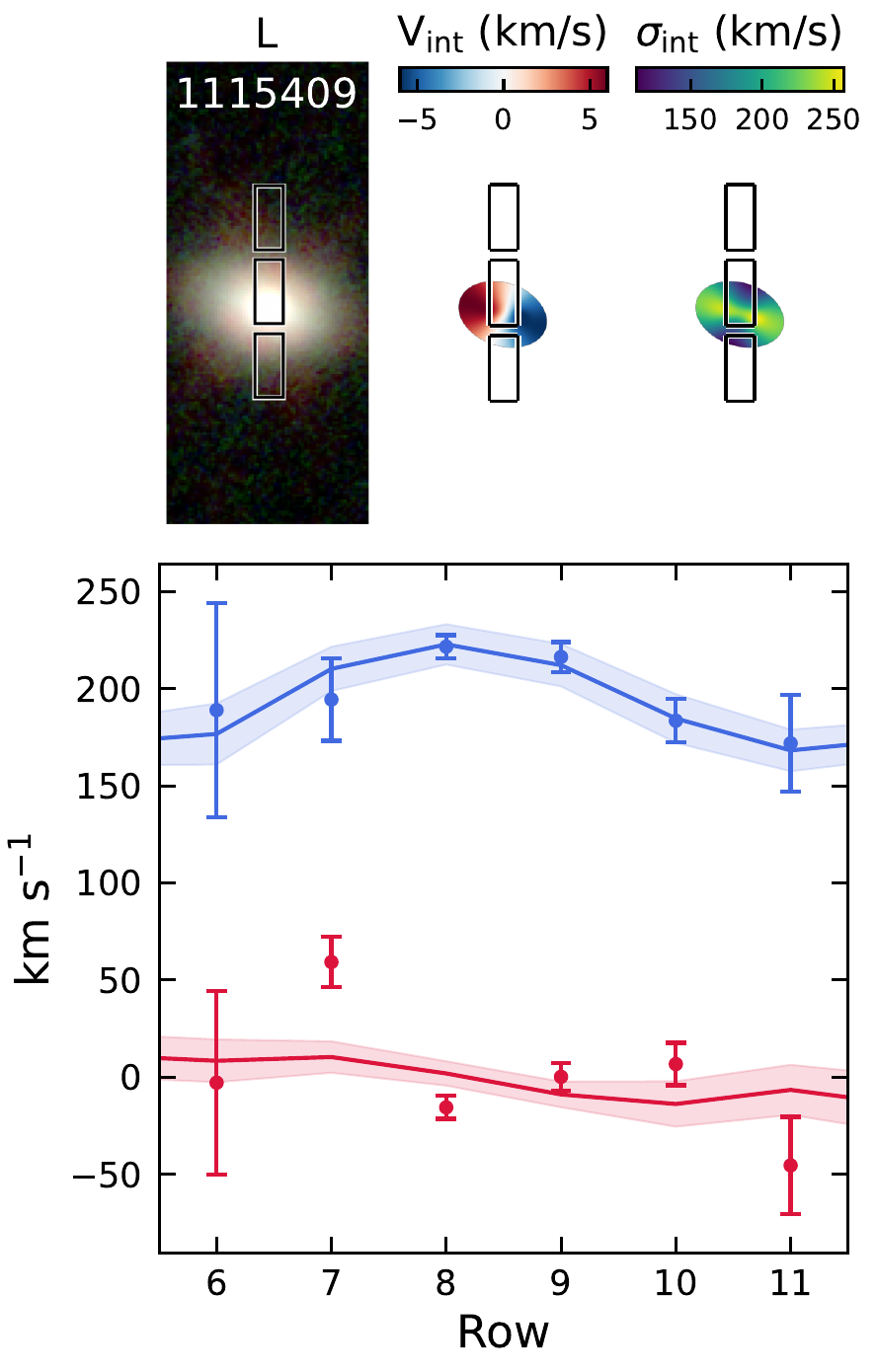}}
    \subfigure{\includegraphics[width=0.30\textwidth]{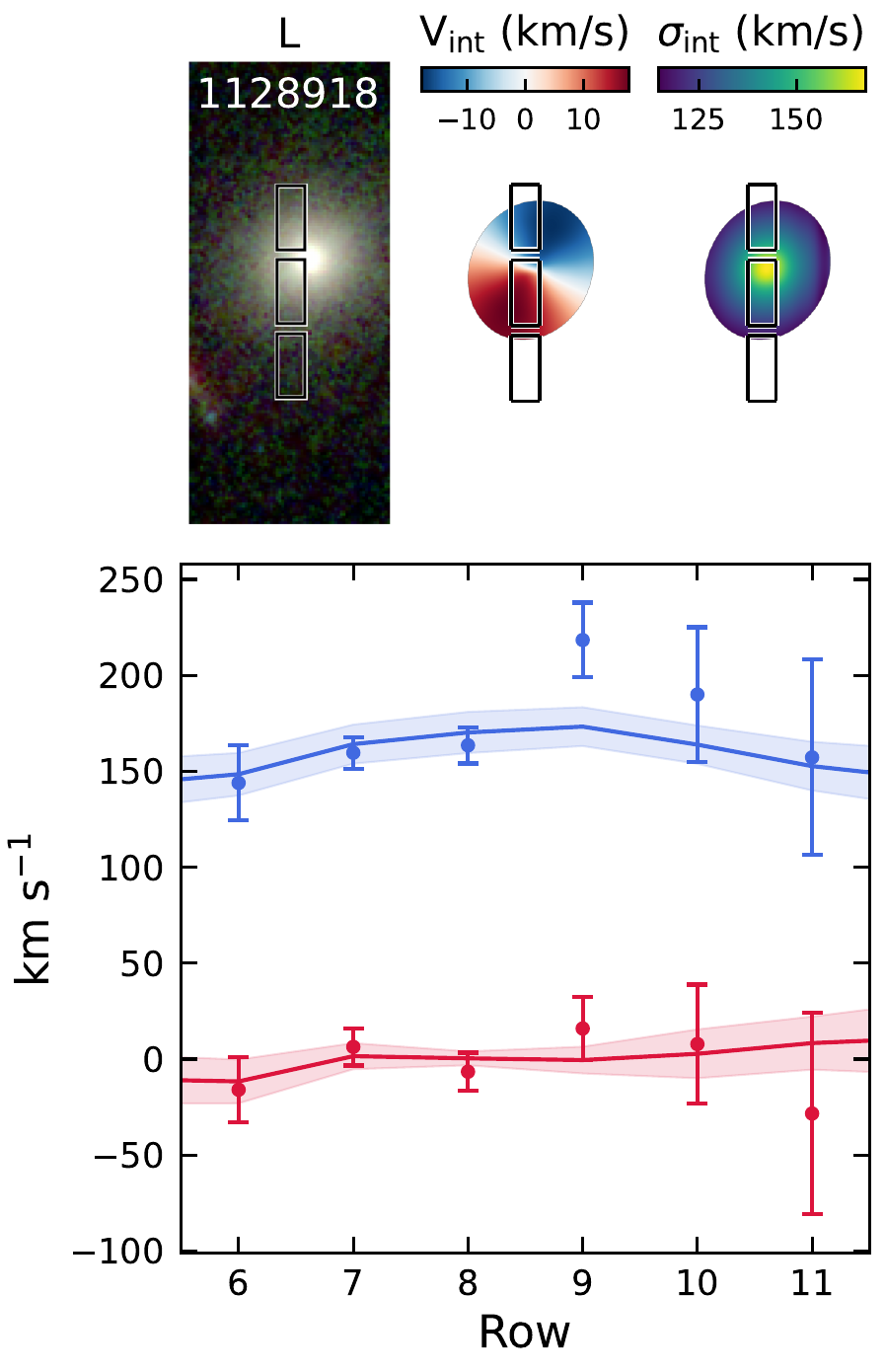}}
    \subfigure{\includegraphics[width=0.30\textwidth]{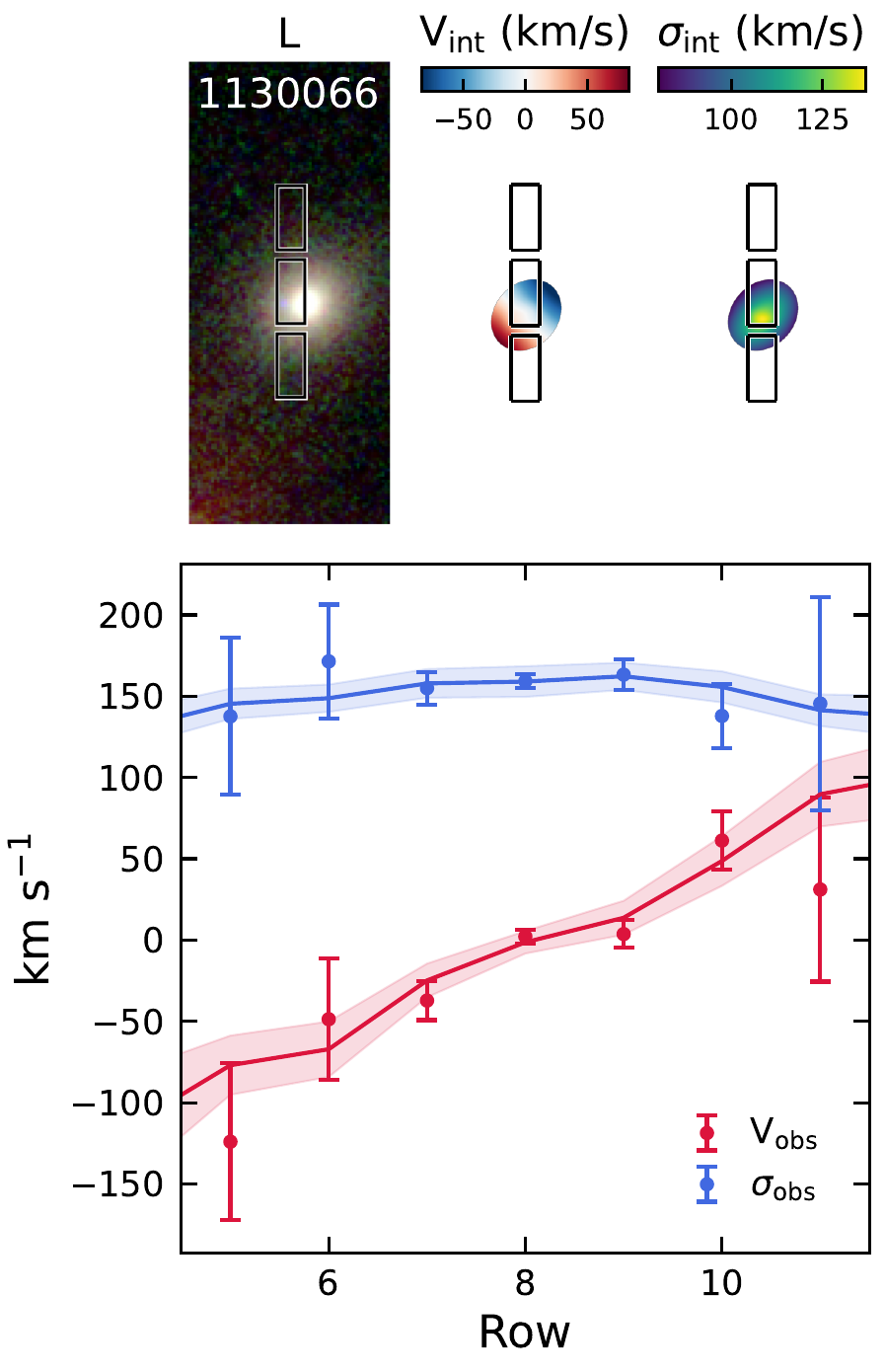}}
    \subfigure{\includegraphics[width=0.30\textwidth]{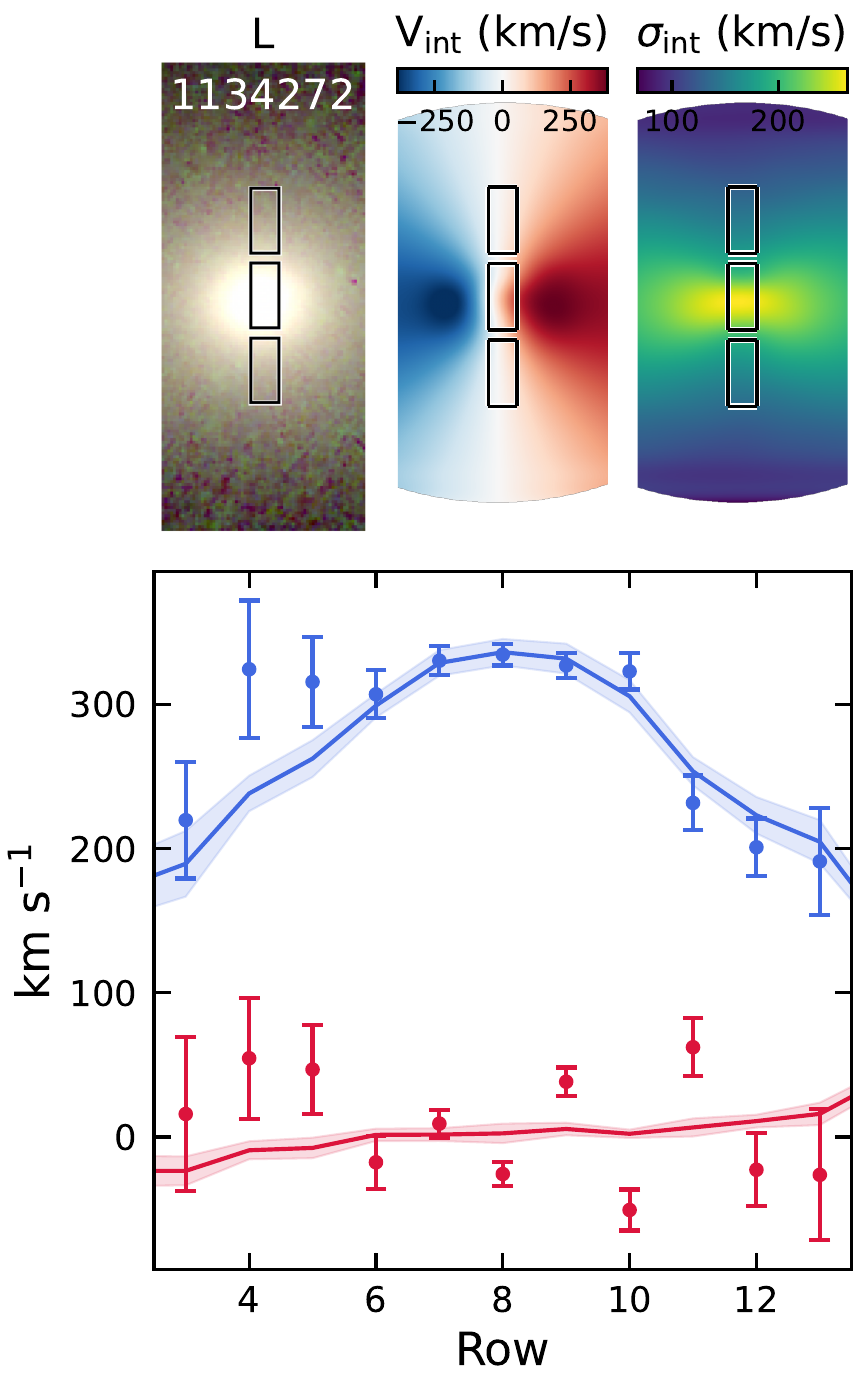}}
    \subfigure{\includegraphics[width=0.30\textwidth]{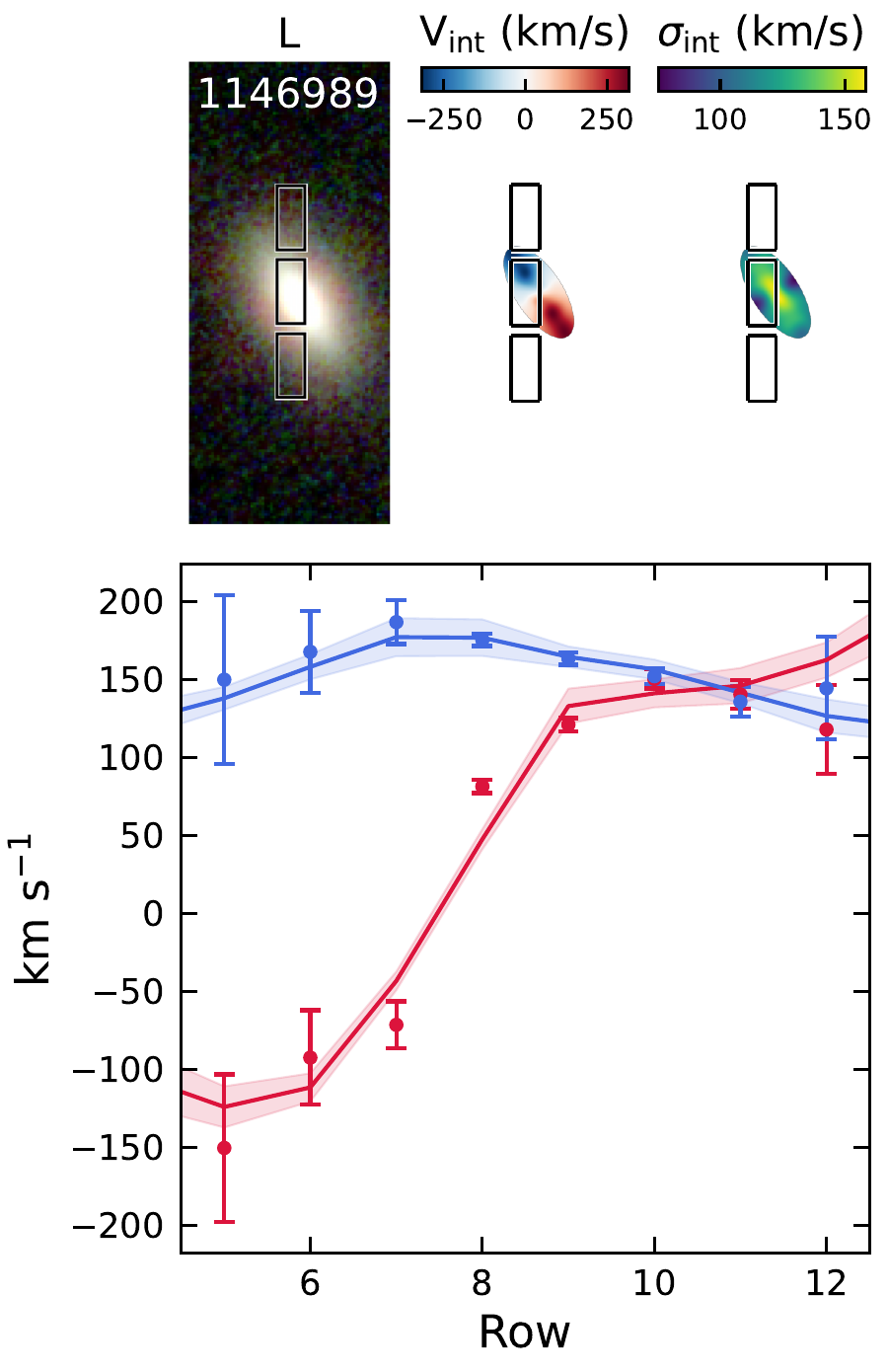}}
    \subfigure{\includegraphics[width=0.30\textwidth]{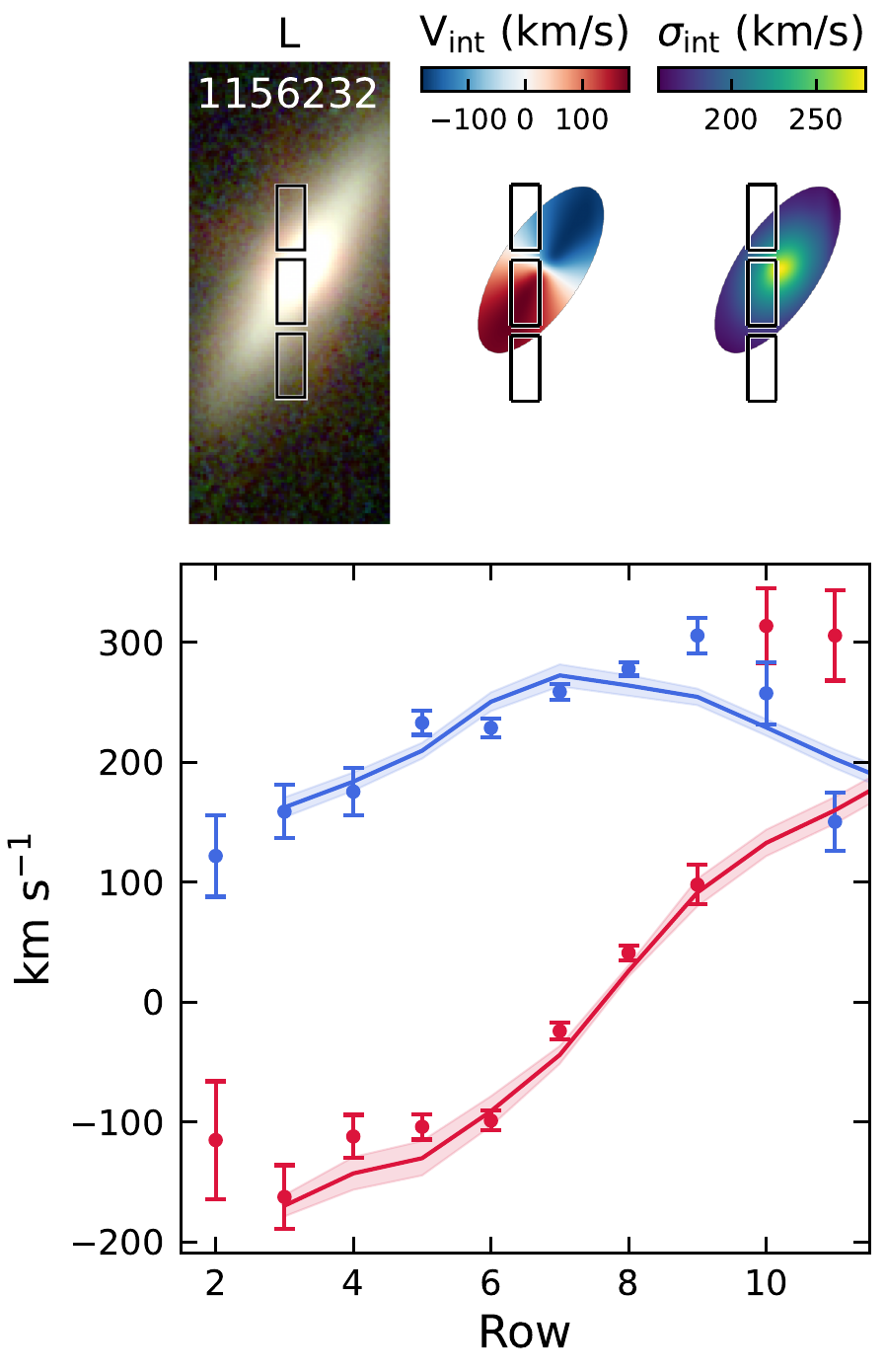}}
    \caption{Observed kinematic profiles and best-fit models for the galaxies in our sample. In the bottom panels we show the 1D observed velocities and velocity dispersion profiles as the red and blue data points, respectively. The 1D best-fit model profiles are shown as the corresponding lines, with the 1$\sigma$ uncertainties indicated by the shaded region around the best-fit profile. In the top panels we show the COSMOS-Web RGB imaging \citep{CCasey2023} (left), and the inferred, unconvolved, 2D line-of-sight velocity (middle) and velocity dispersion (right) profiles out to 2$R_{e}$. We overlay the positions of 3 MSA microshutters on the light, velocity, and velocity dispersion maps.}
    \label{fig:overviews}
\end{figure*}

\begin{figure*}[t]
    \setcounter{figure}{2}
    \centering
    \subfigure{\includegraphics[width=0.30\textwidth]{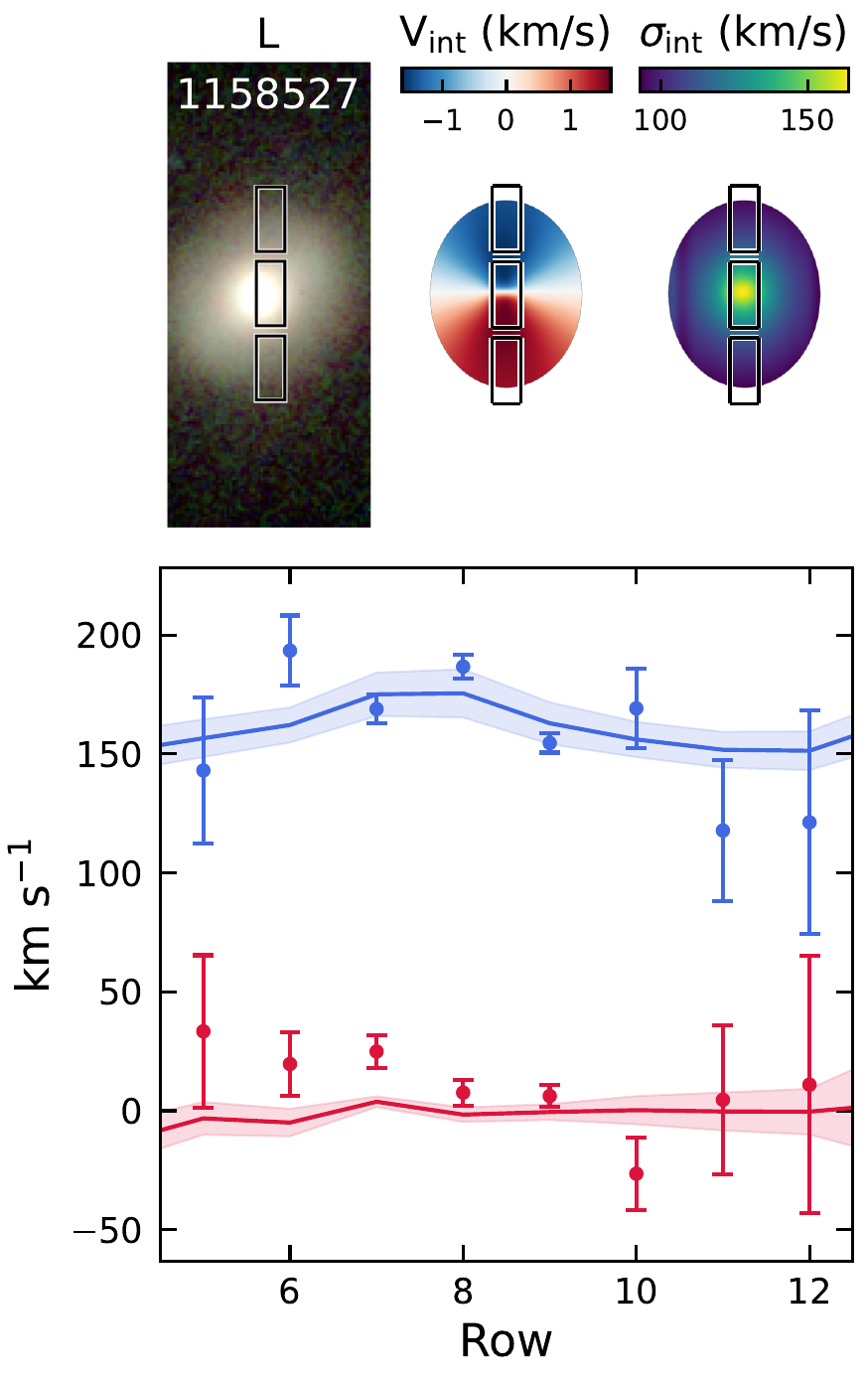}}
    \subfigure{\includegraphics[width=0.30\textwidth]{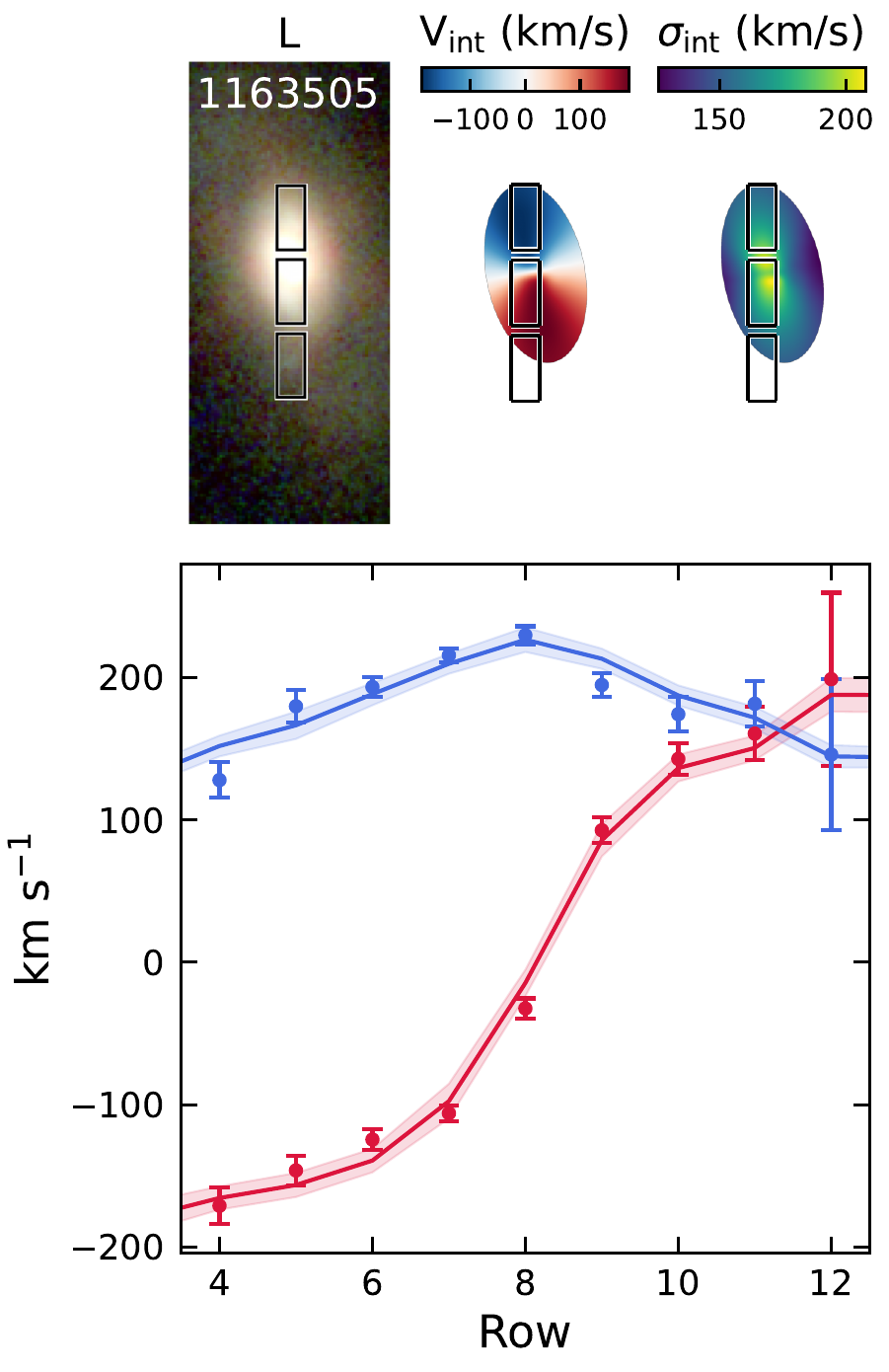}}
    \subfigure{\includegraphics[width=0.30\textwidth]{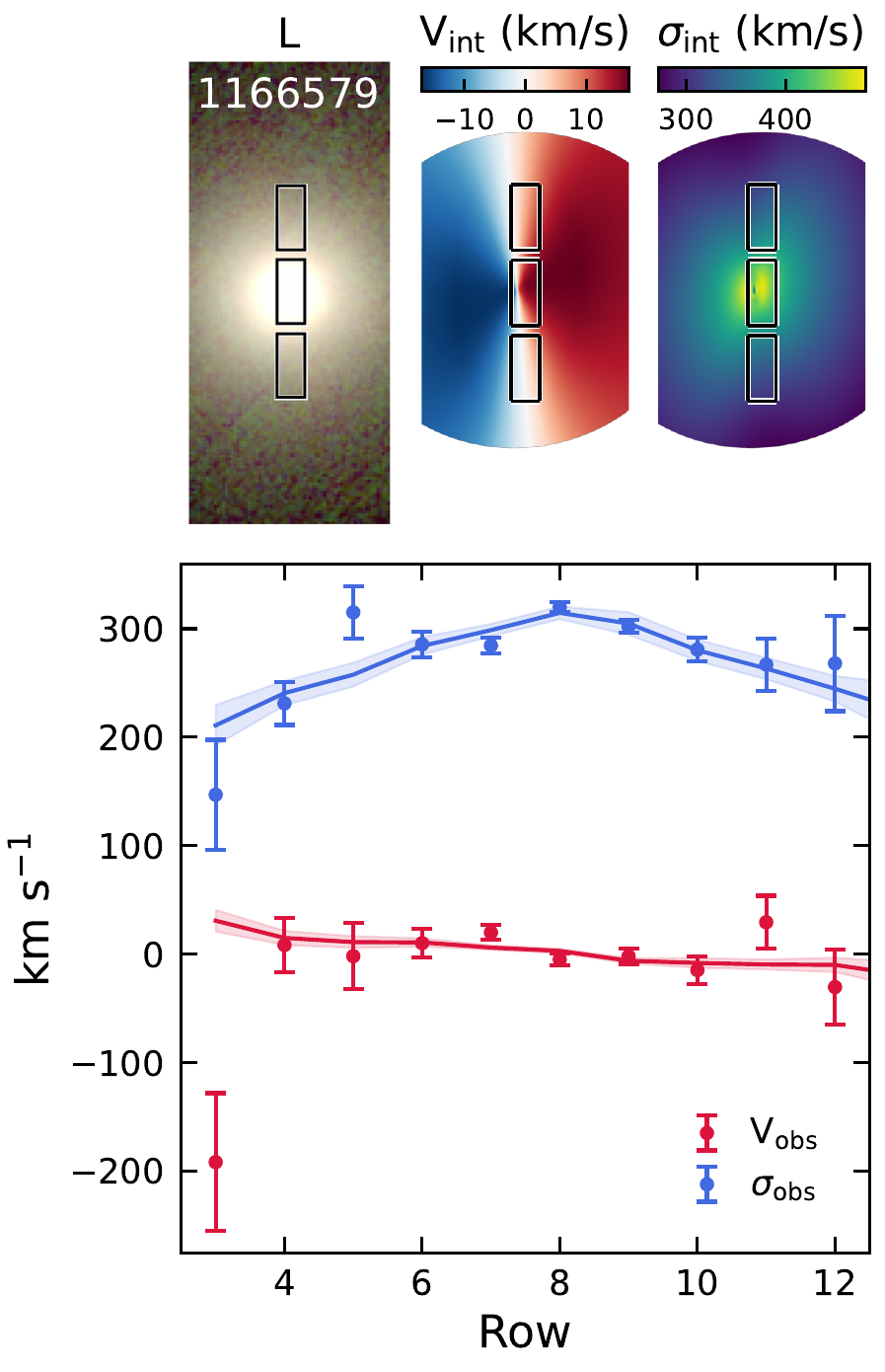}}
    \flushleft
    \caption{Continued.}
\end{figure*}

\subsection{Dynamical masses}\label{sec:res_dyn_masses}
\begin{figure*}
    \centering
    \includegraphics[width=\linewidth]{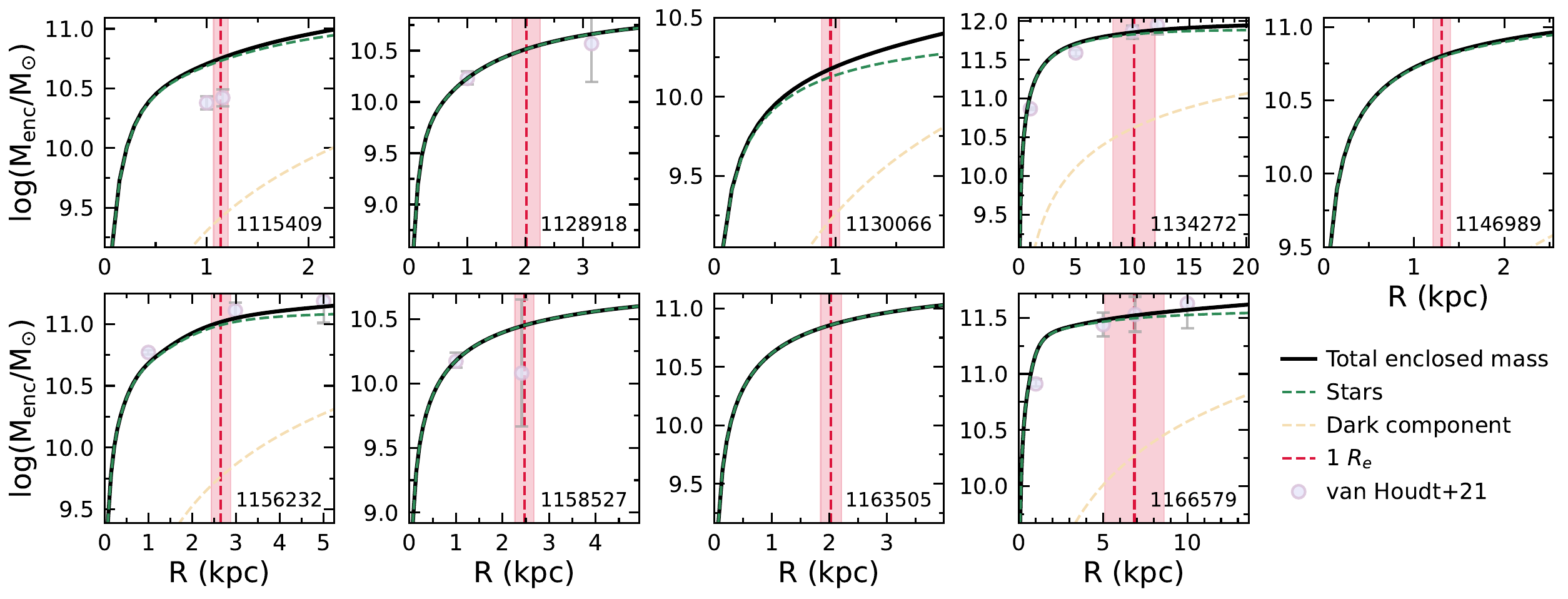}
    \caption{Enclosed total mass as a function of intrinsic radius for the best-fit JAM models of the galaxies in our sample. The red dashed line indicates $1R_e$, the limit used to calculate the JAM masses in this work. The green dashed line indicates the stellar contribution to the mass, and the yellow dashed line indicates the DM contribution. The grey data-points show the enclosed masses from JAM models derived from the VLT/VIMOS LEGA-C spectra \citep{JvanHoudt2021}, where available.}
    \label{fig:mass_growth_curve}
\end{figure*}

We use the best-fit JAM dynamical models to derive the dynamical masses for the galaxies in our sample. To this end we compute the enclosed mass by spherically integrating the gravitational potential that follows from our dynamical model out to 1$R_e$. In Figure \ref{fig:mass_growth_curve} we show the enclosed mass as a function of radius for the galaxies in our sample. 

We also show the enclosed mass profiles derived by \citet{JvanHoudt2021} from LEGA-C spectra, based on much lower spatial resolution (ground-based) VLT/VIMOS spectra, with a nearly perpendicular slit orientation, for six common galaxies as grey circles. Given the $\sim7\times$ smaller PSF FWHM of our observations, our measurements can be used to assess how well these less resolved ground-based measurements are able to recover the spatial structures of the galaxies in our sample. The enclosed mass profiles from these two observations are in good agreement (within 1$\sigma$) for five galaxies, while for the galaxy 1115409 the enclosed mass profiles we derive here are $\sim0.2$~dex higher compared to those derived from LEGA-C observations. We discuss the comparison between the two observations further in Section \ref{sec:disc_caveats}.

The gravitational potential includes both the stellar and an additional ``dark" component. This dark component accounts for variation in the total $M/L$, and could be due to dark matter, $M_\star/L_\star$ variations, and gas contributions (see Section \ref{sec:met_mass_models}). The complete gravitational potential description therefore represents the total mass distribution of the galaxy. However, from Figure \ref{fig:mass_growth_curve} we see that within 1$R_e$ the dark component contributes a negligible fraction of the total mass for all galaxies in our sample.

We obtain the total dynamical mass of each galaxy by multiplying the enclosed mass within $1R_e$ by a factor 2, following the conventional approximation that the stellar half mass radius is equal to the half light radius widely used in studies of galaxy dynamics. Uncertainties on the dynamical masses are derived from the Monte Carlo simulations used to obtain errors on the best-fit models (see Section \ref{sec:met_par_inference}). The resulting JAM dynamical masses are reported in Table \ref{tab:results}.

Given the complexity of our dynamical models, we assess the robustness of the inferred dynamical masses by examining their dependence on the parameters in the model. We find that the dynamical masses are primarily constrained by the stellar mass-to-light ratio and dark matter normalisation, both of which are tied directly to the observed velocity dispersion profiles. The remaining five parameters largely determine the rotational structure and orbital anisotropy, and therefore have a limited effect on the inferred mass and velocity dispersion profiles. Significant degeneracies exist for these parameters given the one-dimensional nature of our data, especially for galaxies with MSA shutters orientated primarily along their minor axis. However, the masses of these galaxies remain robust as they are predominantly constrained by their velocity-dispersion profiles. For galaxies observed along along their major axis the spatially resolved rotational velocity curves provide additional constraints on the dynamical masses. Furthermore, the observed light distribution further restricts the allowed mass models.

Although our data therefore do not permit robust measurements of rotational support or anisotropy for every galaxy, the inferred dynamical masses are well constrained. We verify this using galaxy-by-galaxy recovery tests spanning a range of input masses (Appendix \ref{ap:recovery_test}). Our method recovers these masses with a mean offset of only $-0.016 \pm 0.004$~dex, well within the typical uncertainties on our dynamical mass estimates.\\

\begin{figure}[t]
    \centering
    \includegraphics[width=0.9\linewidth]{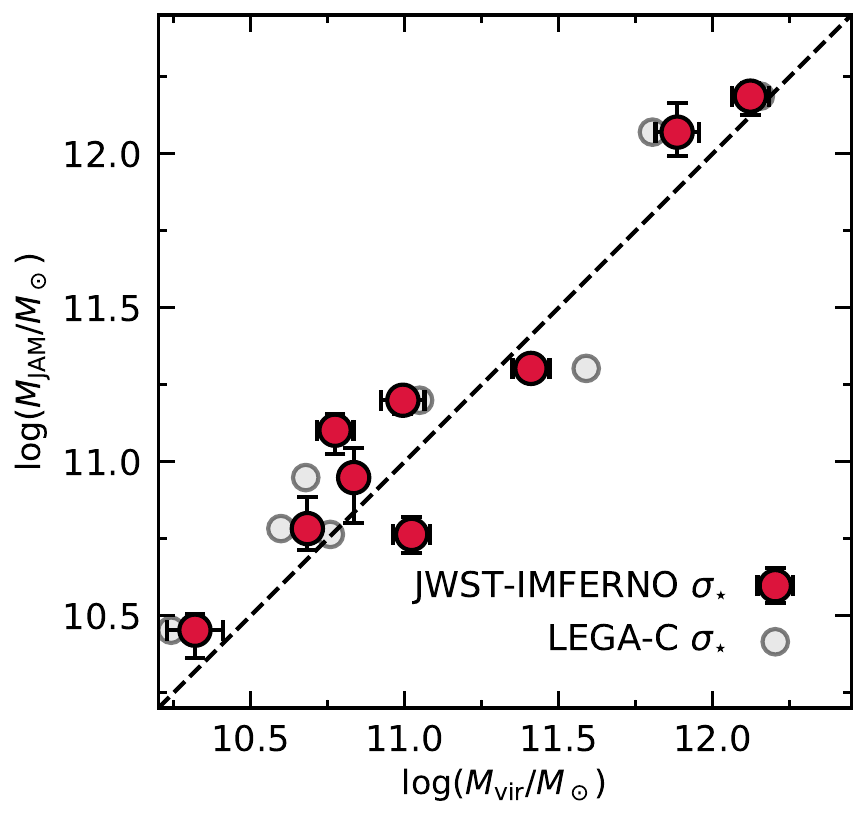}
    \caption{Virial masses versus the JAM dynamical masses derived in this study. The dashed line indicates a one-to-one relation. The red data points show the virial masses that were calculated using the integrated velocity dispersion from the JWST-IMFERNO spectra, while the grey data points show the virial masses calculated using the velocity dispersions from the LEGA-C spectra \citep{AvanderWel2021}.}
    \label{fig:mvir_mjam}
\end{figure}

In Figure \ref{fig:mvir_mjam} we compare the JAM dynamical masses to virial masses to assess systematic offsets between the two methods. Since these virial masses can depend significantly on the aperture and alignment of the data, we calculate these masses using the derived integrated velocity dispersions from both the NIRSpec/MSA JWST-IMFERNO spectra and the LEGA-C spectra of each galaxy. These integrated velocity dispersions were fit using \texttt{alf} by \citet{CCheng2026} for both data sets.
We calculate the virial masses using the F150W half-light radii (see Section \ref{sec:met_mgefit}) and the Sers\'ic correction factor from \citet{MCappellari2006}. We evaluate this correction using the Sers\'ic index of the best-fit single Sers'ic model, as the correction is defined for single-component profiles. Although we use double Sers'ic models throughout the rest of this work, the resulting differences in Sers'ic index have only a minor effect on the inferred virial masses, which falls within the quoted uncertainties.

We show the virial masses derived from the JWST-IMFERNO spectra in red, and the LEGA-C spectra in grey.
For most galaxies, we find that the virial masses calculated using the two different observations are in good agreement with each other, though significant offsets (up to $\sim0.5$\,dex)  exists for individual sources. 

When we adopt the maximum virial mass between the two observations, we find that for seven out of nine galaxies the virial and JAM dynamical masses agree within 2$\sigma$, with a mean offset of $0.10\pm0.20$~dex. The similarity between the two dynamical mass calculations implies that the virial theorem provides a reasonable approximation for the dynamical mass for the galaxies in this sample. For three galaxies in our sample (1115409, 1146989, 1156232) the JAM dynamical masses differ from the virial estimates by $>0.25$~dex. For 1146989 and 1156232 this offset may be due to the fact that these two sources are significantly off-centred in the MSA slit, while for 1115409 the slit is aligned mostly along the minor axis of the galaxy. Under such geometric effects, the integrated velocity dispersion, used to calculate the virial mass, can be significantly biased \citep{MSlob2025}, indicating that caution should be used when interpreting the virial masses of individual galaxies. We discuss the implications of our results for dynamical mass measurements in more detail in Section \ref{sec:disc_MDyn}.\\

\subsection{Constraints on the IMF}\label{sec:res_IMF}
\begin{figure*}[t]
    \centering
    \includegraphics[width=\textwidth]{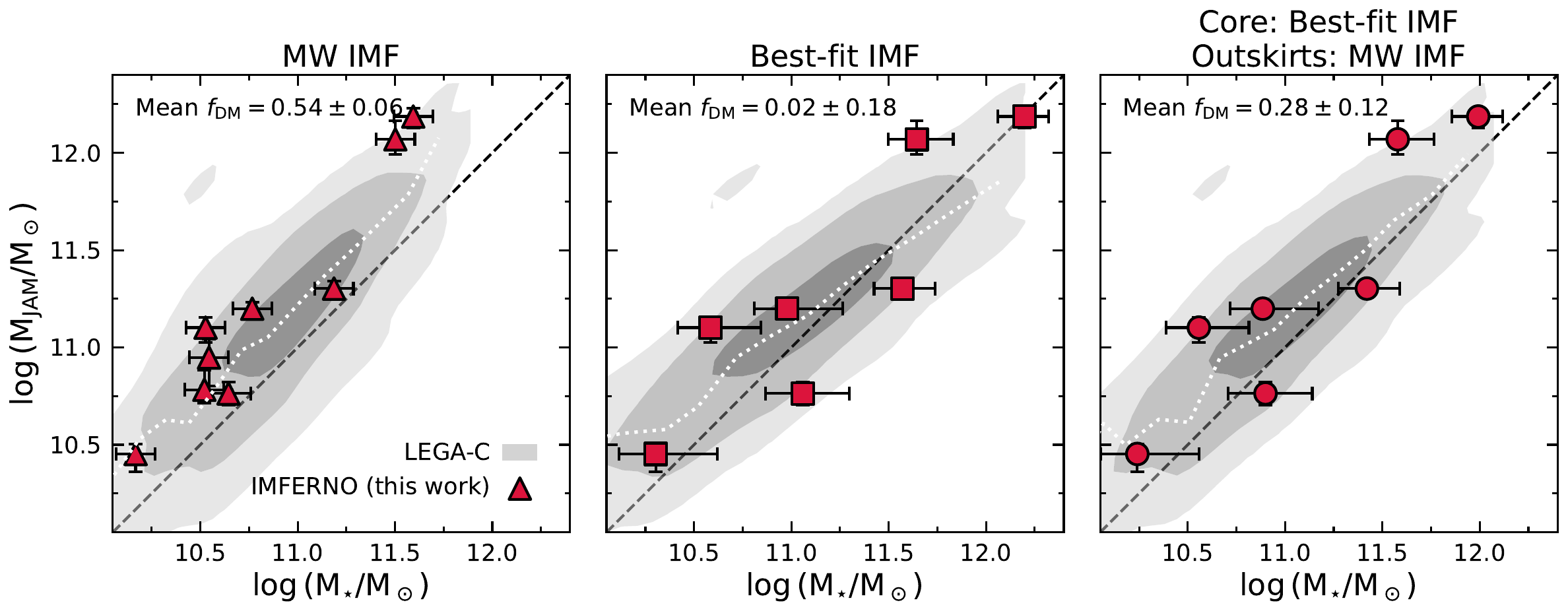}
    \caption{vJAM dynamical masses versus stellar masses, calculated using a \citet{PKroupa2001} IMF (left panel), the best-fit IMF from \citet{CCheng2026} (middle panel), and a radially varying IMF (right panel). The IMFERNO galaxies from our sample are shown as the red data points. The contours show the distribution of all quiescent galaxies in the LEGA-C survey, corrected for the relevant IMF using a relation between IMF mismatch parameter and velocity dispersion from \citet{TTreu2010}. We also show the running mean of the LEGA-C population as the dotted white line. The dashed line indicates a one-to-one relation, where points below this line fall in the unphysical regime. }
    \label{fig:mjam_mstar}
\end{figure*}

With the derived JAM dynamical masses we can now assess whether the stellar masses of our galaxies are consistent with the galaxy's dynamics under different IMF assumptions. First, we show the comparison between the dynamical masses and the \texttt{Prospector} stellar masses for a MW-like \citet{PKroupa2001} IMF in the left panel of Figure \ref{fig:mjam_mstar}.

Under a MW IMF, all galaxies satisfy $M_{\rm dyn} > M_\star$, with a corresponding mean DM fraction within 1~$R_e$ of $f_{\rm DM} = 0.54 \pm 0.06$. This DM fraction was calculated assuming a negligible gas mass, as the galaxies in our sample are old ($> 1$~Gyr; \citealt{CCheng2026}) and we thus expect little remaining gas in these systems \citep{JSpilker2018, bezanson_now_2022, zanella_large_2023,gangula_resolved_2026}. The inferred DM fraction we find is substantially higher than typical measurements of ETGs at $z\sim0$ at similar masses ($f_{\rm DM} \sim 0.2$; \citealt{cappellari_atlas3d_2013,santucci_sami_2022,KZhu2024,yang_dark_2024}), and would require strong decrease in the central DM content between $z\sim0.7$ and $z\sim0$. In contrast, simulations generally predict that the DM fraction within 1$R_e$ increases, albeit slightly, over such timescales \citep{remus_co-evolution_2017, lovell_fraction_2018}, driven by an increase in the effective radius \citep[e.g.][]{JvandeSande2013,AdeGraaff2024}.
Thus, while a Milky Way IMF produces dynamically consistent mass ratios, the implied high DM fractions make this IMF implausible.

Next, we compare our dynamical masses to stellar masses that are corrected for a best-fit IMF. This IMF was derived from gravity sensitive absorption features in the IMFERNO spectra by \citet{CCheng2026} for seven out of nine galaxies in our sample. Three galaxies favour bottom-heavy IMFs, increasing their stellar masses by factors of $\sim2-4$ relative to a MW IMF, while the remaining galaxies are consistent with a MW IMF but still exhibit a modest mass increase of $\sim10-60$\%. These stellar mass corrections are quantified through the IMF mismatch parameter, $\alpha_{\rm IMF}$, which rescales the stellar masses derived using a \citet{PKroupa2001} IMF. In the middle panel of Figure \ref{fig:mjam_mstar} we compare the dynamical masses to the IMF-corrected stellar masses. Adopting the best-fit IMF brings the stellar masses and dynamical masses closer in agreement. Nonetheless, for two galaxies in our sample the stellar masses exceed the dynamical masses within 2$\sigma$, and we find a mean DM fraction of $f_{\rm DM} = 0.02 \pm 0.18$ for this IMF assumption.

While such a low central DM fraction is physically possible, it is implausible that the most massive galaxies in our sample have a negligible amount of DM given their large effective radii \citep{remus_co-evolution_2017,DXu2017,lovell_fraction_2018,AdeGraaff2024}.
This tension may not be surprising, as we applied the best-fit IMF correction to the full galaxies, while the measurements from \citet{CCheng2026} were derived using integrated spectra which are weighted toward the central $R_e$ of the galaxy. Moreover, observations of ETGs at $z\sim0$ suggest that the IMF is bottom-heavy in the central regions of a galaxy, but approaches a MW IMF at larger radii \citep[e.g.][]{IMartinNavarro2015,FLaBarbera2016,TDavis2017,PvanDokkum2017,MKianusch2024}. Applying the \citet{CCheng2026} $\alpha_{\rm IMF}$ mass-correction globally may therefore overestimate the total stellar mass if a gradient was already are in place at $z\sim0.7$. 

We test this scenario in the right panel of Figure \ref{fig:mjam_mstar}, where we adopt a radially varying IMF using a simple step function; we apply the measured $\alpha_{\rm IMF}$ correction within 1$R_e$, and use $\alpha_{\rm IMF} = 1$ beyond 1$R_e$, corresponding to a MW IMF. In practice, this parametrisation corresponds to multiplying half the mass of the galaxy by $\alpha_{\rm IMF}$.
We emphasise that this step function is merely a simplified parametrisation of a potential IMF gradient, as no direct measurements of IMF gradients currently exist at $z\sim0.7$. However, the adopted parametrisation is motivated by observations of local ETGs, where the transition from a bottom-heavy to MW IMF typically occurs at a physical scale comparable to $1R_e$ in our sample ($\sim3$~kpc; \citealt{PvanDokkum2017}). Nonetheless, we note that galaxies for 1134272 and 1166579 1$R_e$ is significantly larger than this physical scale, meaning we potentially overestimate the fraction of stars that have a bottom-heavy IMF for these galaxies. For consistency we choose to correct all galaxies using the same method, but emphasise that future direct measurements of an IMF gradient in these galaxies are needed to properly constrain the stellar masses.

For the model in which we only correct the centre of the galaxy for the best-fit IMF, the stellar masses are more consistent with our dynamical constraints, with a mean $f_{\rm DM} = 0.28 \pm 0.12$ and all galaxies have $M_{\rm dyn} > M_\star$ within the 1$\sigma$ limit. The mean DM fraction we find for this IMF is thus comparable to those of massive ETGs at $z\sim0$ \citep{cappellari_atlas3d_2013,santucci_sami_2022,KZhu2024,yang_dark_2024}. However, we note that the DM fractions we derive in this work are degenerate with the choice in parametrisation of the IMF gradient, and are thus poorly constrained for our arbitrary description of this gradient. It is therefore difficult to directly compare the DM fractions across different epochs without direct measurements of the $M/L$ from the IMF.

Nonetheless, we conclude that the radially varying IMF yields physically plausible DM fractions while remaining broadly consistent with dynamical constraints. We discuss the implications of these results further in Sections \ref{sec:disc_IMF} and \ref{sec:disc_gal_evol}.

\section{Discussion}\label{sec:discussion}
\subsection{Implications for virial masses}\label{sec:disc_MDyn}
In this work we have measured the JAM dynamical masses of nine massive quiescent galaxies at $z\sim0.7$ from NIRSpec/MSA observations using a forward modelling approach. This method accounts for MSA projection effects, slit losses, and potential rotational support, all of which can affect dynamical mass estimates. Interestingly, from Figure \ref{fig:mvir_mjam} we conclude that our JAM dynamical masses are broadly consistent with the dynamical masses that were computed using the virial theorem. A similar result was found for the LEGA-C sample, where JAM masses were based on modelling the VLT/VIMOS LEGA-C spectra \citep{JvanHoudt2021, AvanderWel2022}.

However, we note that for individual galaxies the difference between the JAM and virial dynamical mass measurements can be significant. For the full LEGA-C sample, the mean scatter between the two methods is 0.15~dex, with deviations reaching up to 0.4~dex \citep{AvanderWel2022}. Similarly, in our sample, three out of seven galaxies show significant offsets of $>0.25$~dex between the two estimates. We also emphasise that the errors on the virial masses are likely underestimated, as they only capture statistical errors from the data, but do not include the systematic uncertainties involved in calculating virial masses using different methods \citep{AvanderWel2022}.

Comparing the virial masses calculated with the JWST-IMFERNO integrated velocity dispersions to virial masses calculated with the LEGA-C integrated velocity dispersions shows that these observations yield virial masses that are significantly offset for individual galaxies. This result likely reflects the fact that the slit apertures from NIRSpec/MSA are much smaller than those of LEGA-C, such that different regions of the galaxies are probed in the two observations. Additionally, the significantly lower spatial resolution of the LEGA-C observations could explain the offset in virial masses.  Furthermore, the almost perpendicular orientations of the NIRSpec and LEGA-C slits can also explain the offset in virial masses, as different kinematic orientations are probed with the two slits. While aperture corrections have been established to mitigate such slit losses and alignment effects \citep{JvandeSande2013, SPrice2022}, defining such a correction for the often significantly mis-aligned and offset slits of NIRSpec/MSA observations, in particular for rotating galaxies, is non-trivial \citep{MSlob2025}. Additionally, by not applying an aperture correction to the velocity dispersion, while using the total effective radius and Sers\'ic index, the quantities used to calculate the virial mass do not self-consistently probe the same regions of the galaxy. These challenges underscore the need for the detailed forward-modelling approach presented in this work to obtain reliable dynamical masses from NIRSpec/MSA studies.

\subsection{Extending IMF constraints to LEGA-C}\label{sec:disc_IMF}
In Section \ref{sec:res_IMF} and Figure \ref{fig:mjam_mstar} we presented how the dynamical masses of the galaxies in our sample compare to their stellar masses under different IMF assumptions. Such a comparison provides an independent check on the IMF mismatch parameter measured by \citet{CCheng2026}, as the dynamical masses do not rely on stellar population modelling assumptions. 
We find that a MW IMF yields dynamical masses that are consistent with the stellar masses, although systematically high compared to plausible central DM fractions. If we instead adopt the best-fit IMF measured by \citet{CCheng2026} for the galaxies in our sample, the stellar masses significantly exceed the dynamical masses for two out of seven galaxies, and the mean inferred DM fraction is low ($\sim0$). Applying the \citet{CCheng2026} best-fit IMF only to the inner 1$R_e$, and using a MW IMF beyond, represents an intermediate case, reducing the inferred DM fractions compared to a MW IMF, while keeping the stellar masses consistent with our dynamical constraints.

To test this IMF description for a larger sample of galaxies, we also include 569 quiescent galaxies from the LEGA-C sample \citep{AvanderWel2021} in Figure \ref{fig:mjam_mstar} as grey contours, adopting virial masses as dynamical mass estimates. While these virial masses are less reliable than the JAM masses we derive in this work, given the overall agreement of our dynamical and virial masses (see Section \ref{sec:res_dyn_masses}), we expect that using the virial masses rather than JAM masses for the LEGA-C sample will not systematically affect our results.
For the uniformly corrected IMF case, we correct the stellar masses using an empirical relation between $\alpha_{\rm IMF}$ and the stellar integrated velocity dispersion, $\sigma_{\star}$, derived from the dynamics of $z\sim0$ ETGs by \citet{TTreu2010}. This trend is consistent with the $\alpha_{\rm IMF}-\sigma_\star$ relation found for the JWST-IMFERNO sample \citep{CCheng2026}, and thus a valid assumption for the general population of galaxies at $z\sim0.7$ (see also \citealt{JMendel2020}).
For the core-only bottom-heavy IMF model we apply the same step-function prescription we adopted for the IMFERNO sample (see Section \ref{sec:res_IMF}), using the $\alpha_{\rm IMF}-\sigma_{\star}$ relation in the central 1$R_e$ of the galaxy, and a MW IMF beyond.

As expected, these (partially) bottom-heavy IMF-corrections shift the LEGA-C galaxies to systematically higher stellar masses compared to the MW IMF. The corresponding mean DM fractions within 1$R_e$ are $f_{\rm DM} = 0.32 \pm 0.34$ for the MW IMF, $f_{\rm DM} = 0.18 \pm 0.39$ for the uniformly corrected IMF, and $f_{\rm DM} = 0.25 \pm 0.33$ for the core-only corrected IMF. The mean central DM fractions are thus not problematic, and do not differ significantly, for the three assumed IMF models. This smaller change in mean DM fraction assumptions for the LEGA-C sample compared to the IMFERNO sample can be explained by the fact that the LEGA-C population has a lower average velocity dispersion, and thus a less strong IMF correction compared to IMFERNO. However, for the LEGA-C sample the most massive galaxies ($\log_{10}M_\star/M_\odot > 11$) enter the unphysical regime where $M_\star > M_{\rm dyn}$ for the fully bottom-heavy IMF, suggesting that their stellar masses are overestimated for this IMF. 
However, when we limit the IMF correction to the central regions for the LEGA-C sample, the population mean falls within the physical regime for all stellar masses. Furthermore, only the radially varying IMF model produces stellar and dynamical masses that remain consistent with a slope of 1, whereas the MW and free IMF assumptions lead to systematic positive and negative offsets at the high-mass end, respectively. Such a constant relation between stellar and dynamical masses has also been found for observations of $z\sim0$ ETGs with the IMF as a free parameter, and simulations at intermediate redshift, which find no strong dependence of the DM fraction on stellar mass \citep{remus_co-evolution_2017, busillo_casco_2025}.

Overall, the radially varying IMF yields the most uniform trend over the full mass range for both the JWST-IMFERNO and LEGA-C samples. This outcome is consistent with studies of local massive ETGs, where bottom-heavy IMFs are primarily confined to galaxy centres \citep{IMartinNavarro2015,FLaBarbera2016,TDavis2017,PvanDokkum2017,MKianusch2024}. Although the precise form of any IMF gradient remains unconstrained at these redshifts, our results favour a spatially varying IMF that is most bottom-heavy toward the galaxy centres probed by our NIRSpec/MSA data.

\subsection{Implications for galaxy evolution}\label{sec:disc_gal_evol}
As discussed in the previous section, for the most massive galaxies at $z\sim0.7$ the dynamical masses are most consistent with the stellar masses when we correct the central regions of the galaxies for the bottom-heavy IMF from \citet{CCheng2026}. Such bottom-heavy cores are also measured in local ETGs \citep{IMartinNavarro2015,FLaBarbera2016,TDavis2017,PvanDokkum2017,MKianusch2024, MGu2026}. Our findings therefore suggest that these cores were already in place by $z\sim0.7$, and support a scenario in which galaxy growth between $z\sim0.7$ and $z\sim0$ occurs predominantly in the outskirts of galaxies, likely through minor merger-driven growth, while the central regions remain largely undisturbed \citep[e.g.][]{RBezanson2009,SPatel2013,PvanDokkum2010, KSuess2020}. This scenario is reinforced by the fact that undisturbed `relic' galaxies at $z\sim0$ are also found to have a bottom-heavy IMF at all radii \citep{IMartinNavarro2015, martin-navarro_inspire_2023, maksymowicz-maciata_inspire_2024}. Furthermore, the similarity in age and [Fe/H] gradients of massive quiescent galaxies at $z\sim0.7$ and $z\sim0$ \citep{Cheng_2024, JGreene2015, IMartinNavarro2018} indicate little evolution in stellar population gradients over this timescale.

Extrapolating this scenario to higher redshifts, the bottom-heavy cores of the most massive $z\sim0.7$ JWST-IMFERNO galaxies are the plausible descendants of massive quiescent galaxies at earlier epochs \citep{CCheng2026}. In this context it is interesting to note that all galaxies in our sample reside in an overdense environment consisting of multiple groups at different stages of virialisation \citep{BDarvish2017, iovino_high_2016} (see also Fig. \ref{fig:sample_overview}), which are themselves likely falling together into a larger structure. Such an overdense environment is expected to promote accelerated growth and earlier quenching compared to field galaxies \citep[e.g.][]{stanford_evolution_1998,DThomas2005,muzzin_gemini_2012, zavala_gas_2019,ditrani_cosmos_2025}, consistent with these systems being massive and quenched at early cosmic times.
We also note that the most massive $z>3$ quiescent galaxies observed preferentially reside in overdense regions \citep{KIto2023,mcconachie_where_2025}. Together, these results suggest that both our sample and their potential high-redshift progenitors share a common environmental context.

With this in mind, one galaxy in our sample has an inferred formation redshift $z_{\rm form} \gtrsim 5$ (ID: 1134272 \citealt{CCheng2026}), making it a likely descendant of the $z>4$ quenched population. Notably, this galaxy is also the most bottom-heavy galaxy in our sample ($\alpha_{\rm IMF} = 4.01_{-0.85}^{+0.71}$). However, we do not rule out that the other galaxies in our sample also descend from the population of high-$z$ quiescent galaxies, as significant mergers or late time star formation may have polluted their stellar populations, leading to younger inferred luminosity-weighted ages. Within a minor merger-driven growth framework, in which mass is primarily added to the outskirts, this result suggests that the bottom-heavy IMF observed in galaxy centres at $z\sim0.7$ was already in place in quiescent galaxies at $z>4$, potentially increasing their inferred stellar masses by a factor $\sim3$.

Detailed dynamical modelling will enable us to assess the IMF beyond $z\sim0.7$ in more detail in the future. First, deep, spatially resolved IFU observations from stellar kinematics will allow us to place more stringent constraints on the mass distribution, including dark matter, and directly measure $M_{\rm dyn}/L$ gradients across individual galaxies. Furthermore, applying the methods presented in this work to larger samples of $z>1$ quiescent galaxies will yield accurate dynamical masses, which we can use to independently confirm their stellar masses and constrain the IMF. Such measurements will clarify whether a bottom-heavy IMF was already in place in massive quiescent galaxies at $z>0.7$, and whether these systems evolve into the cores of local ETGs through minor-merger accretion in their outskirts or if late-time star formation plays a more disruptive role.

\subsection{Caveats}\label{sec:disc_caveats}
In this work we have derived the dynamical masses of nine massive quiescent galaxies from the JWST-IMFERNO program using a forward-modelling approach applied to resolved NIRSpec/MSA spectra. In our sample we have made no selection on the alignment between the galaxy position angle (PA) and the MSA shutter. As a result, for two galaxies the slit is orientated primarily along the minor axis ($\Delta$PA$>60$\textdegree). For these galaxies, the observed velocity data provides limited information on the intrinsic rotational structure, and the resulting velocity fields are difficult to interpret. We therefore restrict our analysis to the inferred dynamical masses, which are driven mainly by the observed velocity dispersion for these galaxies, and are therefore less sensitive to these geometric effects. 

For six of the galaxies in our sample \citet{JvanHoudt2021} previously derived Jeans models (see Figure \ref{fig:mass_growth_curve}). Their measurements are based on HST/F814W imaging and lower spatial resolution VLT/VIMOS LEGA-C spectra, obtained with slits oriented almost perpendicular to our NIRSpec/MSA shutters. In Appendix \ref{ap:legac_comparison} we show the spatial resolution obtained from our data (PSF FWHM $\sim0.1\arcsec$) compared to LEGA-C observations (PSF FWHM $\sim0.75\arcsec$). Despite the different observing geometries and significant improvement in spatial resolution, we find good overall agreement with the inferred mass profiles of the LEGA-C observations.

For one galaxy (ID 1115409), however, our enclosed mass is $\sim0.2$~dex higher than those derived by \citet{JvanHoudt2021}. This galaxy has the NIRSpec/MSA slits oriented close to its minor axis, suggesting that slit misalignment may introduce additional uncertainty in the inferred dynamical masses. However, galaxies 1134272 and 1166579 are similarly misaligned and show good agreement with the LEGA-C mass profile, so any discrepancy between the two methods likely cannot be attributed to slit orientation alone. Instead, the discrepancy between the two enclosed mass profiles of 1115409 could also be explained by the small linear size of the galaxy ($R_e = 1.14$~kpc), which is unresolved in the LEGA-C observations of \citet{JvanHoudt2021}. The remaining compact galaxies in our sample (IDs 1130066 and 1146989) did not yield reliable measurements to be included in the \citet{JvanHoudt2021} sample. This result indicates that, especially for compact galaxies, the improved spatial resolution of NIRSpec/MSA is needed to measure resolved stellar dynamics and obtain reliable dynamical masses.

In addition to the orientation with respect to the major axis, the projection of the galaxies could be another limitation.
Several galaxies in our sample are close to round in projection (i.e. have high axis ratios), making their intrinsic PAs uncertain. To test the impact of this uncertainty, we repeated the fits with the PA as a free parameter and found no significant change in the derived dynamical masses. Nonetheless, a more complete assessment of systematic uncertainties associated with slit misalignment would require comparison with deep, spatially resolved IFU observations, which are not currently available for these galaxies.

Our mass models include a dark component which effectively captures radial variation in the mass-to-light ratio (see Section \ref{sec:met_mass_models}). Such a variation may arise from dark matter, gas, or stellar population gradients. The gas contribution is expected to be negligible for these old galaxies \citep{JSpilker2018, bezanson_now_2022, zanella_large_2023}, while \citet{Cheng_2024} find only mild stellar population gradients in LEGA-C galaxies. The fitted component is therefore most likely sensitive to dark matter and $M/L$ gradients, which cannot be reliably disentangled given the limited spatial resolution and radial extent of our data. Moreover, we fix the slope of the DM profile to a NFW profile, while studies of lensed galaxies suggest that DM profile are shallower beyond $z\sim0.5$ \citep{WSheu2025}. We therefore do not interpret the best-fit dark matter parameters or inferred $M/L$ gradients. Nevertheless, the stellar component dominates the enclosed mass within 1$R_e$ (see Fig. \ref{fig:mass_growth_curve}), indicating that our total dynamical masses are only weakly sensitive to the adopted parametrisation of the dark component.

Finally, we note that we derive the total JAM dynamical masses from the mass enclosed within the half-light radius, rather than the half mass radius. This approach provides a consistent definition across our dynamical models. However, radial variations in the stellar mass-to-light ratio can cause the half-mass radius to differ from the half-light radius \citep{KSuess2019a}. Consequently, converting the mass within the half-light radius to a total mass may introduce a modest systematic uncertainty, whose magnitude depends on the strength of the underlying $M_*/L$ gradient. Future spatially resolved constraints on stellar-population and IMF gradients can be used to quantify this effect, and obtain the total mass self-consistently.

\section{Conclusions}\label{sec:conclusion}
We present JAM dynamical masses of nine massive quiescent galaxies at $z\sim0.7$ from the JWST-IMFERNO survey. For the most massive galaxies in this sample \citet{CCheng2026} have measured an excess of low-mass stars compared to a MW IMF using dwarf-sensitive absorption features. In this paper we assess whether these IMF measurements are supported by dynamical constraints.

The galaxies in this work were selected from the LEGA-C spectroscopic program \citep{AvanderWel2021}, and observed for $\sim$32 hours in a single NIRSpec/MSA pointing. The resulting ultra-deep spectra provide spatially and spectrally resolved stellar absorption features in rest-frame $\sim0.6-1.1$~$\mu$m, from which we measure the resolved stellar velocity and velocity dispersion profiles along the $2-4$\arcsec slitlets composed of $5-8$ MSA shutters.

To interpret these kinematic profiles, which were measured with randomly orientated and offset MSA slits compared to the galaxy's major axes, we developed a detailed forward-modelling framework. In this approach we use Jeans modelling using JAM \citep{MCappellari2008} to model the intrinsic kinematics. This dynamical model is based on the light distributions derived from JWST/NIRCam imaging from the COSMOS-Web program \citep{CCasey2023}. Next, we generated mock observations of these models with \texttt{msafit} \citep{AdeGraaff2023}, taking into account the slit geometry, galaxy morphology, instrument optics, and observational and data reduction effects. The best-fitting JAM kinematic profiles directly yield a model of the gravitational potential, from which we derive dynamical masses.

When we compare the JAM dynamical masses to stellar masses calculated with a MW IMF, we find they are consistent, but result in a relatively high DM fraction ($f_{\rm DM, 1R_e}=0.54\pm0.06$) compared to local ETGs of similar stellar masses. If we instead apply the IMF which was measured by \citet{CCheng2026} for seven galaxies in our sample, parametrised using the IMF mismatch parameter ($\alpha_{\rm IMF}$), we find a mean DM fraction that is significantly lower ($f_{\rm DM, 1R_e}=0.02\pm0.018$) compared to $z\sim0$ ETGs. Furthermore, with this IMF we find that for two out of seven galaxies the stellar masses exceed the dynamical masses by more than 2$\sigma$, suggesting the masses are overestimated. When we only apply the IMF correction to the inner 1$R_e$, and use a MW IMF beyond 1$R_e$, the stellar masses of all galaxies in our sample are consistent with the dynamics within 1$\sigma$. Such a radially varying IMF is similarly found in $z\sim0$ ETGs. For this radially varying IMF model we find a mean $f_{\rm DM, 1R_e}=0.28\pm0.12$, similar to nearby ETGs. Nonetheless, we note the exact dark matter fractions of the galaxies in our sample are not well constrained without an exact parametrisation of their radial IMF gradients. Overall, our work supports the bottom-heavy IMF found by \citet{CCheng2026}, but we conclude that the $\alpha_{\rm IMF}$ measurements likely only represent the central regions of these galaxies.

We also compare the inferred JAM dynamical masses with the virial masses, and find that for most galaxies in our sample they agree well. However, for three galaxies the offset is $>0.25$~dex, highlighting that virial masses of individual objects can be unreliable.

Despite the offset between virial and JAM mass for individual galaxies, the average agreement enables us to assess the IMF for the full LEGA-C sample using their virial masses. For the MW IMF we find that the LEGA-C sample occupies the same region of the $M_\star-M_{\rm dyn}$ plane as the JWST-IMFERNO sample. When we correct the stellar masses for the IMF, using the empirical relation between $\alpha_{\rm IMF}$ and the stellar velocity dispersion $\sigma_\star$ derived by \citet{TTreu2010}, we find that the most massive galaxies in the LEGA-C sample shift into the unphysical regime where $M_\star > M_{\rm dyn}$. However, when we only correct for the IMF within 1$R_e$, similar as for the JWST-IMFERNO sample, galaxies of all stellar masses are consistent with their dynamical constraints, roughly following a one-to-one relation. 
To confirm whether this radially varying IMF is indeed in place at $z\sim0.7$, future studies, either directly measuring the IMF at different apertures of $z\sim0.7$, or using dynamical constraints, are needed.

Our results imply that the bottom-heavy cores of $z\sim0$ ETGs were already in place at $z\sim0.7$. This finding is consistent with a scenario in which galaxy growth at later times occurs primarily at large radii, while the central regions remain largely unchanged, likely through minor merger-driven assembly. Given the old age and very early formation time ($z_{\rm form} > 5$) of our most bottom-heavy IMFERNO target, we expect that massive quiescent galaxies at much earlier times also have an excess of low-mass stars compared to the MW IMF. Such a bottom-heavy IMF would increase their stellar masses up to a factor of 4, making their masses even harder to reconcile with current models of galaxy formation \citep{JAntwiDanso2025,AWeibel2025, AdeGraaff2024_highz_q}. Future studies measuring the IMF at earlier epochs are required to confirm this scenario, either through gravity sensitive absorption features or dynamical constraints.

Finally, this study demonstrates that detailed forward modelling enables robust dynamical mass measurements for distant massive quiescent galaxies using NIRSpec/MSA data, even in the presence of severe slit losses and random slit orientations and placements. Such measurements are essential for assessing the IMF and validating stellar masses inferred from SPS modelling in high-$z$ systems. Thus, deep NIRSpec/MSA programs combined with forward-modelling methods provide an efficient avenue for building statistical samples of distant quiescent galaxies with dynamical measurements.

\begin{acknowledgements}
This work is based on observations made with the NASA/ESA/CSA James Webb Space Telescope. The data were obtained from the Mikulski Archive for Space Telescopes at the Space Telescope Science Institute, which is operated by the Association of Universities for Research in Astronomy, Inc., under NASA contract NAS 5-03127 for JWST. These observations are associated with program JWST-GO-05629. Support for program JWST-GO-5629 was provided by NASA through a grant from the Space Telescope Science Institute, which is operated by the Association of Universities for Research in Astronomy, Inc., under NASA contract NAS 5-03127. The specific observations analysed can be accessed via \href{https://doi.org/10.17909/mxwj-am85}{doi: 10.17909/mxwj-am85}.
We thank the COSMOS-Web and LEGA-C teams for making their data publicly available.
Some of the data products presented herein were retrieved from the Dawn JWST Archive (DJA). DJA is an initiative of the Cosmic Dawn Center (DAWN), which is funded by the Danish National Research Foundation under grant DNRF140.
MK acknowledges funding from the Dutch Research Council (NWO) through the award of the Vici grant VI.C.222.047.
AdG acknowledges support from a Clay Fellowship awarded by the Smithsonian Astrophysical Observatory. 
\end{acknowledgements}

\bibliographystyle{aa}
\bibliography{mybib}

\begin{appendix}
\section{GALFIT fitting procedure}\label{ap:galfit}
As described in Section \ref{sec:met_mgefit}, we use GALFIT \citep{CPeng2002} to measure the morphological parameters and construct PSF-deconvolved morphological models of the galaxies in our sample from JWST/NIRCam imaging in the F115W, F150W, F277W and F444W filters \citep{CCasey2023}. Below, we describe our fitting method.

We use the COSMOS-Web images available on MAST\footnote{\url{https://mast.stsci.edu/}} for each filter, cut out to $800\times800$ and $400\times400$ pixel regions centred on each galaxy for the short wavelength (F115W, F150W) and long wavelength (F277W, F444W) filters, respectively. Next, we use \texttt{photutils} to obtain segmentation maps of each cutout to identify neighbouring sources. Before fitting, we mask out any sources more than 2\arcsec\ away from the target, as well as sources that are at least 2.5~mag fainter than the target. For any bright sources within 2\arcsec\ of the target we include an additional component in our fit to model them simultaneously. Finally, we subtract a local background estimate from the masked cutouts.

\begin{figure}
    \centering
    \includegraphics[width=0.85\linewidth]{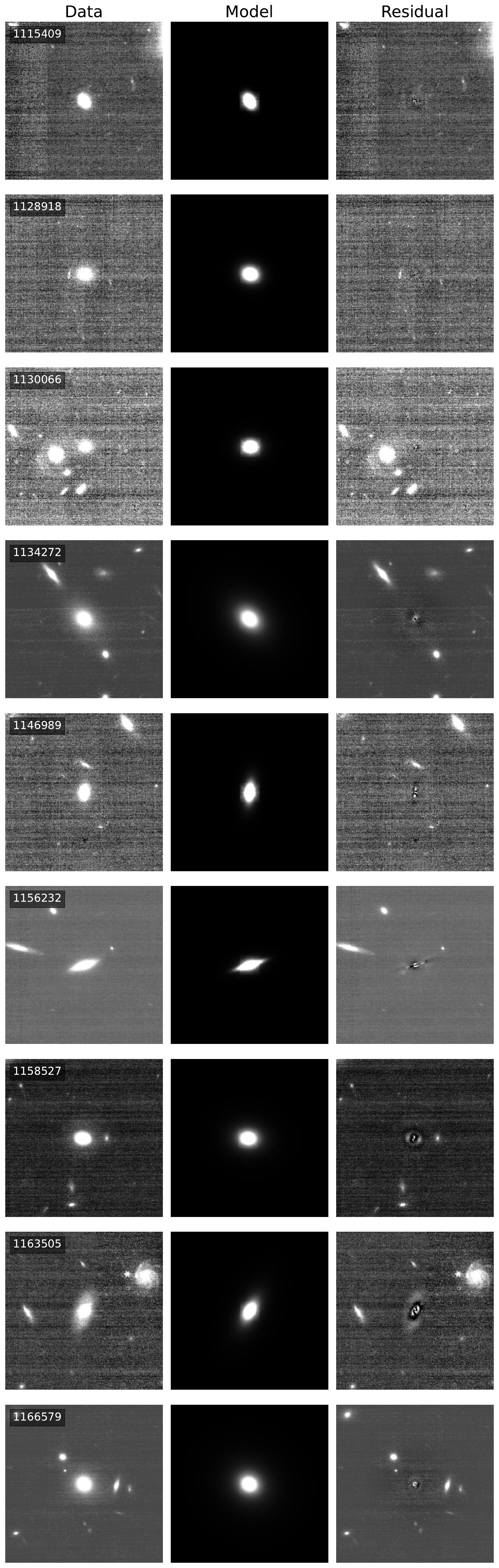}
    \caption{Best-fit \texttt{GALFIT} models F150W for the galaxies in the sample. We show the image cutouts (left), best-fit models (middle), and residuals (right).}
    \label{fig:galfit_fits}
\end{figure}

We use \texttt{STPSF} \citep{MPerrin2014} to construct distortion-corrected PSF models at the native pixel scales of each NIRCam filter. We fit the galaxies in our sample with a double S\'ersic profile, using the PSF model and background-subtracted image as inputs for GALFIT, and derive the structural parameters used throughout this work from the combined profiles. For the half-light radius this corresponds to the radius for which half the light is enclosed by the total light profile from the two S\'ersic models, while for the total axis ratio this corresponds to the light-weighted average of the axis ratios of the two S\'ersic profile ($q_{\rm tot} = (q_1 L_1 + q_2L_2)/(L_1 + L_2)$. In Figure \ref{fig:galfit_fits} we show the images, GALFIT best-fit models, and residuals of the galaxies our sample.

While we adopt a two-component Sers\'ic model in this work, we note that additional Sers\'ic profiles result in smaller residuals for individual galaxies. However, when we propagate these best-fit light models, consisting of three Sers\'ic profiles, through our forward modelling procedure the resulting dynamical masses closely agree (within 1$\sigma$) with those derived using the two component light profile. We thus conclude that our dynamical masses are robust under the assumed light models. We also tested how our best-fit GALFIT light models change under different PSF and convolution box sizes in our GALFIT modelling set-up, and found that the best-fit parameters agree within 1$\sigma$ over a large range of PSF and convolution box sizes.

\section{Priors for \texttt{prospector} fitting}\label{ap:prospector}
As described in Section \ref{sec:met_stellar_masses}, we use \texttt{prospector} \citep{BJohnson2021} to fit the SEDs and spectra of the galaxies in our sample to measure the stellar masses. In Table \ref{tab:prosp_params} we describe the priors for each parameter we use in our fitting procedure.
\begin{table}[h]
    \centering
    \renewcommand{\arraystretch}{1.35}
    \caption{Free parameters and priors for the spectro-photometric modelling with \texttt{Prospector}.}
    \begin{tabular}{p{0.15\textwidth} p{0.15\textwidth} p{0.1\textwidth}}
        \hline\hline
        Parameter & Prior & Limits \\\hline
        $\sigma_\star$ & Uniform & (5, 600) \\
        $\log(Z_\star/Z_{\odot})$ & Uniform & (-1.4, 0.19) \\
        $\hat{\tau}_{\lambda,2}$ & Uniform & (0.0, 2.5) \\
        $\hat{\tau}_{\lambda,1}$ & Fixed to $\hat{\tau}_{\lambda,2}$ & -- \\
        $n_{\rm dust}$ & Uniform & (-1.0, 0.4) \\
        $t$ & Uniform & (0.03, $t_{\rm univ}$) \\
        $\tau$ & Uniform & (0.1, 30) \\
        $\log(M_\star/M_{\odot})$ & Uniform & (10, 13) \\
        $j_{\rm spec}$ & Uniform & (1.0, 5.0) \\
        $f_{\rm outlier}$ & Uniform & (0, 0.5) \\\hline\hline\\
    \end{tabular}
    \label{tab:prosp_params}
\end{table}

\section{Posteriors}\label{ap:posteriors}
\begin{figure*}
    \centering
    \includegraphics[width=\linewidth]{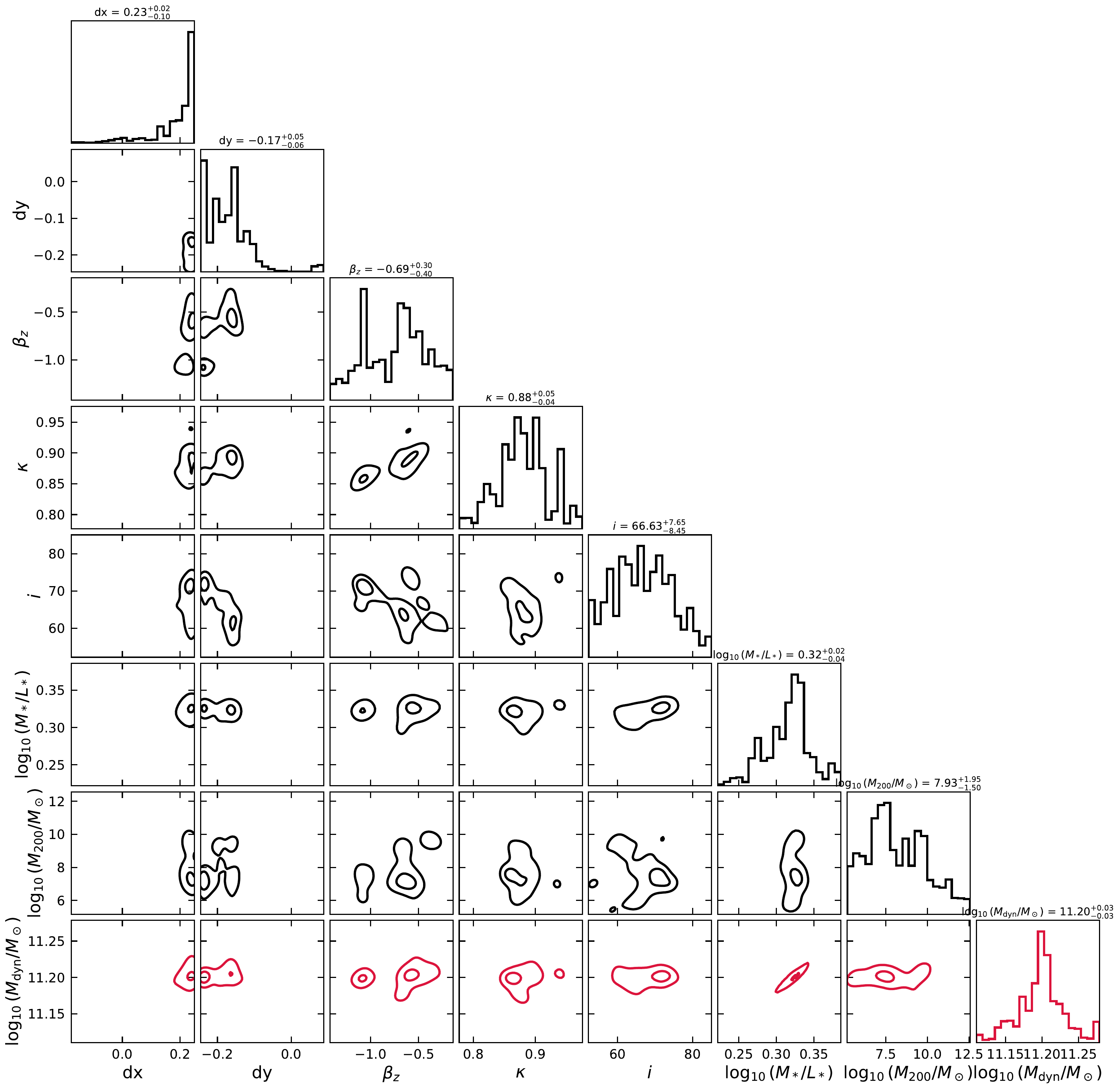}
    \caption{Example posterior for a representative galaxy in our sample (ID: 1163505). The black 1 and 2 $\sigma$ contours indicate parameters that we fit for in our sampling, while the red contour represents the posterior of the inferred dynamical mass resulting from our sampling.}
    \label{fig:ap-posterior}
\end{figure*}

As described in Section \ref{sec:met_par_inference}, we sample over parameters in our fitting procedure. In Figure \ref{fig:ap-posterior} we show the resulting posterior distributions for these parameters, as well as the inferred dynamical mass, for a representative galaxy in our sample (ID: 1163505). The posteriors for the remaining galaxies are available in the online journal. 

\section{Dynamical mass recovery test}\label{ap:recovery_test}
We assess how well our sampling methods recover the dynamical masses by generating a model velocity and velocity dispersion profile using our model pipeline. We generate these profiles for 5 different values of $M_*/L_*$ and $\log(M_{200}/M_{\odot})$, spaced evenly in log space, and leave all other parameters fixed to their best-fit values from our sampling. 
Since the $M_*/L_*$ and $\log(M_{200}/M_{\odot})$ are the main parameters that constrains the dynamical mass, this effectively results in a set of mock observations for different dynamical masses. We also fix the light model to the best-fit light model of the galaxy we are testing. To test how the the galaxy morphology and geometry relative to the slit affects the dynamical recovery, we make these mock observations for all galaxies in our sample. Furthermore, since each galaxy has different best-fit values for the remaining parameters in our fitting procedure, by testing all targets we also assess whether the recovery rate depends on the value of the parameters in the fit.

\begin{figure}
    \centering
    \includegraphics[width=\linewidth]{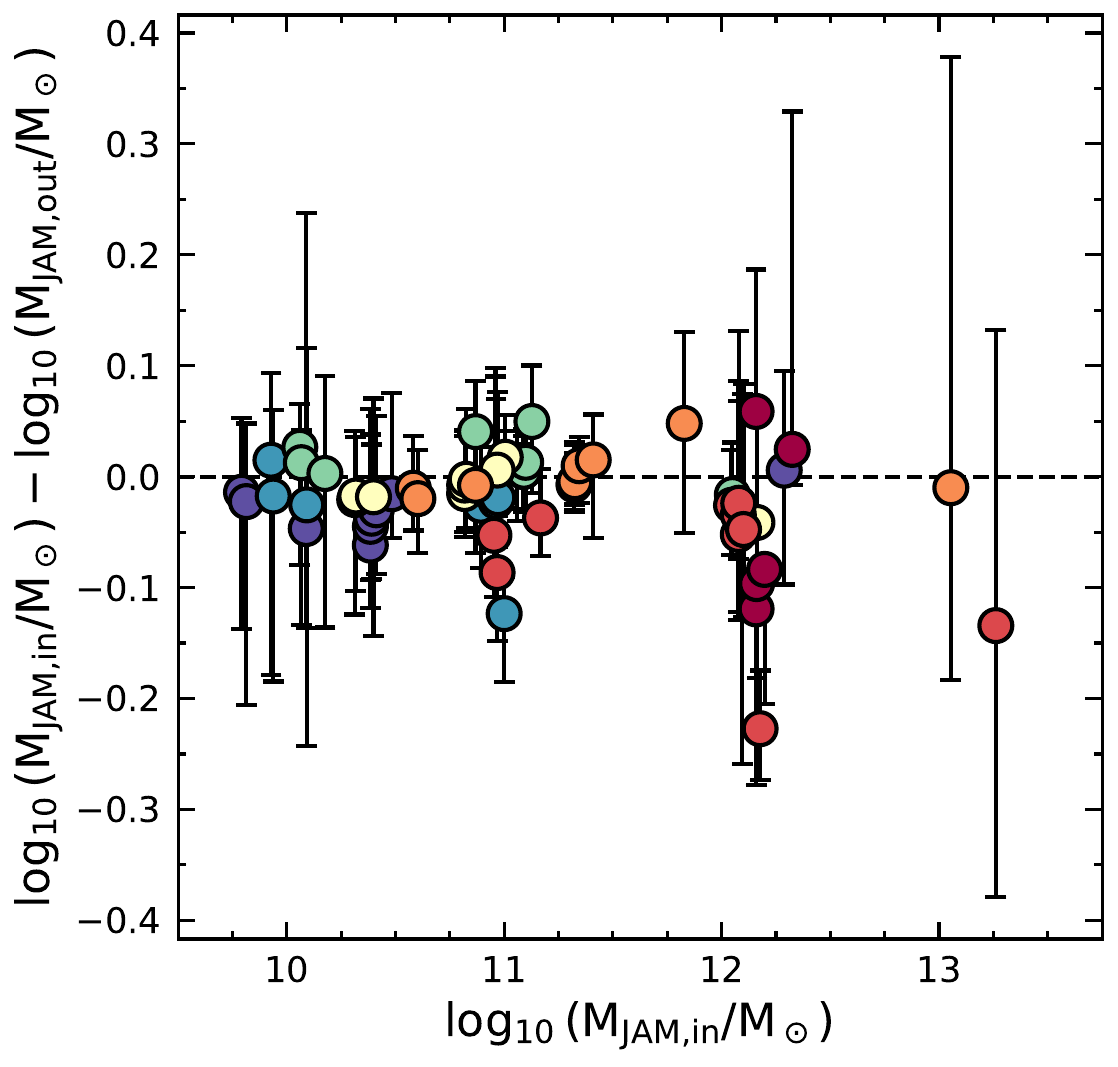}
    \caption{Offset between the input dynamical mass of the mock model and the recovered best-fit dynamical mass, as a function of the input dynamical mass. The data points are coloured for each galaxy using the same colouring convention as Figure \ref{fig:sample_overview}. The dashed line indicates an equal input and recovered dynamical mass.}
    \label{fig:recovery_test}
\end{figure}

We fit these mock velocity and velocity dispersion profiles using the model and sampling set-up described in Section \ref{sec:methods}, and calculate the dynamical masses of the best-fit model to the mock observations. In Figure \ref{fig:recovery_test} we compare the input and recovered dynamical masses for these fits. The points are coloured for each galaxy in the sample using the same colouring convention as Figure \ref{fig:sample_overview}. We find that our model recovers the input dynamical mass within 1$\sigma$ across dynamical masses and for all galaxies in the sample, with a mean offset of $-0.016 \pm 0.004$~dex. This offset is smaller than the typical error on the inferred best-fit dynamical masses we find in this work. 

\section{Comparison to LEGA-C}\label{ap:legac_comparison}
In Section \ref{sec:results} we presented best-fit 2D dynamical models for the galaxies in our sample, derived from resolved slit kinematics of JWST/NIRspec MSA slit kinematics. The resulting dynamical mass profiles are in good agreement with those inferred from ground-based VLT/VIMOS LEGA-C slit observations \citep{JvanHoudt2021} (see Figure \ref{fig:mass_growth_curve}). Here, we illustrate the differences in constraining power between the two observations.

To this end, we generate mock LEGA-C velocity and velocity dispersion fields from our inferred best-fit 2D kinematic models. We convolve the 2D light, velocity, and velocity dispersion profiles with the VLT/VIMOS seeing kernel derived individually for each galaxy by \citet{Cheng_2024}. These kernels are described by Moffat profiles and have a typical FWHM of $\sim0.75\arcsec$ for the galaxies in our sample.

\begin{figure}[t]
    \centering
    \includegraphics[width=0.50\linewidth]{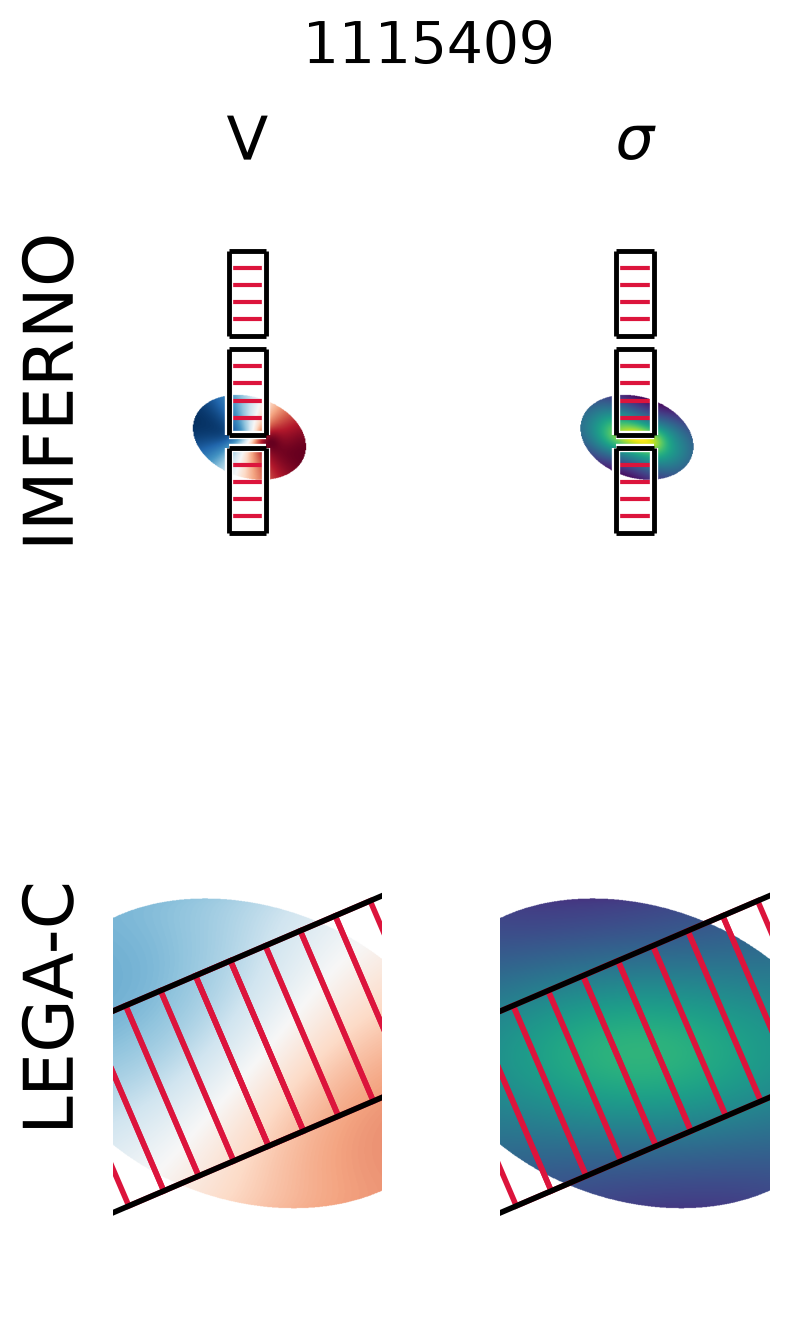}
    \includegraphics[width=0.44\linewidth]{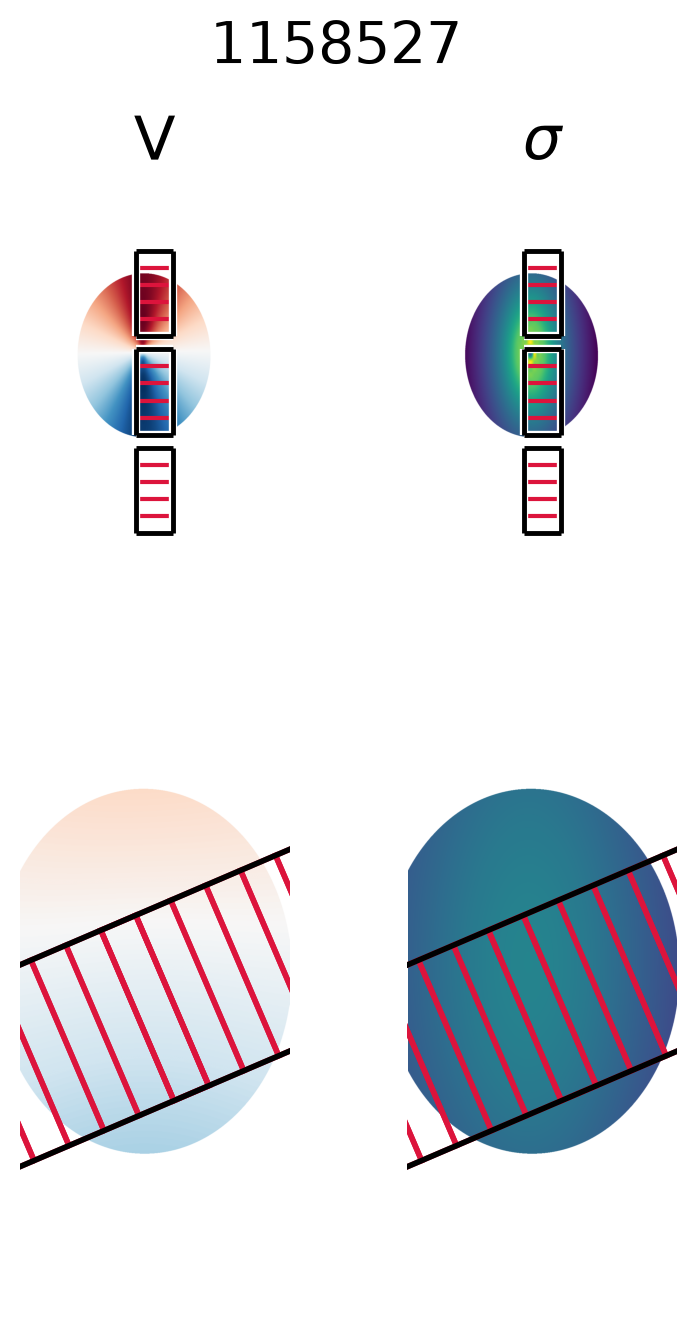}
    \caption{Comparison of the PSF-convolved line-of-sight velocity ($V$) and velocity dispersion fields ($\sigma$) for NIRSpec/MSA (top panels) and VLT/VIMOS (bottom panels) observations for two galaxies in the sample. The fields are generated from the best fit models derived in Section \ref{sec:results}, and convolved with the PSF. The black rectangles in the panels show the slit apertures of the two observations, with the red lines indicating the pixel size. We show the models out to 2 PSF convolved effective radii for both observations.}
    \label{fig:legac_comparison}
\end{figure}

The bottom panels in Figure \ref{fig:legac_comparison} show the resulting PSF-convolved 2D velocity and velocity dispersion maps for two representative galaxies (IDs 1115409 and 1166579). For comparison, the corresponding 2D maps for the JWST-IMFERNO observations are also shown in Figure \ref{fig:legac_comparison}. Here, we convolve the JWST-IMFERNO models with a Gaussian kernel with a FWHM of $0.1\arcsec$, which provides a rough approximation of the PSF of NIRSpec/MSA. 1115409 shows the PSF smearing effect for the most compact galaxy in our sample, while 1158527 shows what the effect is for a galaxy with a size representative of the general sample.

The 2D maps demonstrate the substantial improvement in spatial resolution provided by NIRSpec/MSA. In particular, the JWST observations retain significantly more information on the spatial variation of the velocity dispersion, whereas this structure is largely washed out by the broader VLT/VIMOS PSF. 

We also plot the shutters (slits) of the NIRSpec/MSA (LEGA-C) observations in Figure \ref{fig:legac_comparison}. While the LEGA-C slits are larger compared to NIRSpec/MSA shutters, the large PSF results in much larger convolved galaxy radii, such that these larger slits provide little additional information at extended radii.

\end{appendix}

\end{document}